\documentclass[pra,twocolumn,superscriptaddress]{revtex4}
\usepackage{amsfonts,amsmath,amssymb,float}
\usepackage{bm,dsfont}
\usepackage{color}
\usepackage{bbm}
\usepackage{amsmath,amssymb,lscape,float}
\usepackage{hyperref}
\usepackage{listings}
\usepackage{lipsum}
\usepackage{diagbox}

\usepackage{mathtools}

\newcommand*\widebar[1]{%
  \begingroup
  \def\mathaccent##1##2{%
    \relax
    \ifmmode
      \mkern1.5mu\overline{\mkern-1.5mu#1\mkern-1.5mu}%
      \mkern1.5mu
    \else
      \overline{#1}%
    \fi
  }%
  \overline{#1}%
  \endgroup
}

\usepackage{graphicx}
\usepackage{epstopdf}
\usepackage{dcolumn}
\usepackage{bm}
\usepackage{amsfonts,amsmath,amssymb
}
\usepackage{todonotes}
\usepackage{braket}
\usepackage{verbatim}
\usepackage{siunitx}
\usepackage{xcolor}

\renewcommand{\vec}[1]{\mathbf{#1}}
\renewcommand{\vec}[1]{\ensuremath{\boldsymbol #1 }}

\newcommand{\mrm}{\mathrm}

\DeclareSIUnit\betaz{\beta_\mrm{z}}
\DeclareSIUnit\Er{\mrm{E_{\textrm{R}}}}
\DeclareSIUnit\SystemEnergy{\hbar\mrm{\omega_z}}
\DeclareSIUnit\SystemTime{\mrm{\omega_z}^{-1}}

\begin{document}
\title{Enhanced shortcuts to adiabaticity for coherent atom transport in a family of two-dimensional dynamical optical lattices}

\author{Sascha H. Hauck} 
\affiliation{Department of Flow and Material Simulation, Fraunhofer ITWM, 67663 Kaiserslautern, Germany}
\affiliation{Chair for Scientific Computing, University of Kaiserslautern-Landau (RPTU), 67663 Kaiserslautern, Germany}

\author{Vladimir M. Stojanovi\'c} 
\affiliation{Institut f\"{u}r Angewandte Physik, Technical
University of Darmstadt, 64289 Darmstadt, Germany}

\date{\today}
\begin{abstract}
In view of the compelling need for coherent atom transport as a prerequisite for a variety of emerging quantum technologies, we investigate such transport on the example of an adjustable family of two-dimensional optical lattices [L. Tarruell {\em et al.}, Nature (London) {\bf 483}, 302 (2012)] that includes square, honeycomb, dimerized, and 1D-chains lattices as its special cases; dynamical optical lattices of this type have already been utilized for the demonstration of topological pumping and the realization of two-qubit quantum gates with neutral atoms. At the outset, we propose the appropriate arrangements of acousto-optic modulators that give rise to a frequency imbalance between counterpropagating laser beams, thus leading to the dynamical-lattice effect in an arbitrary direction in the lattice plane. We subsequently obtain the dynamical-lattice trajectories that enable atom transport in the lattices under consideration using two classes of control schemes: (i) shortcuts to adiabaticity (STA) in the form of inverse engineering based on a dynamical
invariant of Lewis-Riesenfeld type, and (ii) their modification, known as enhanced STA (eSTA), which is well-suited for the treatment of anharmonic trapping potentials. We then quantify the resulting atom dynamics using transport fidelities computed from the numerical solutions of the relevant time-dependent Schr\"{o}dinger equations. By doing so for various choices of the system parameters and transport directions, we demonstrate that -- except in the special case of the dimerized lattice -- the eSTA method consistently outperforms its STA counterpart, both in terms of the achievable transport times and the robustness of the resulting transport against small variations of optical-lattice depths.
\end{abstract}

\maketitle
\section{Introduction} \label{Intro}
Recent developments in the realm of emerging quantum technologies -- such as quantum sensing 
based on atom interferometry~\cite{Navez+:16,Dupont-Nivet+:16,Rodriguez-Prieto+:20,Premawardhana+:24,Martinez-Garaot+:25} and 
neutral-atom quantum computing (QC)~\cite{Jaksch+:99,Briegel+:00,
Weitenberg+:11,MohrJensen+:19,Nemirovsky+Sagi:21,Kiefer+:25,
ShiREVIEW:22} -- as well as those in the area of quantum-state engineering~\cite{XFShi:20,
StojanovicPRA:21,Haase+:21,Haase+:22,
Nauth+Stojanovic:22,Stefanatos+Paspalakis:18,
Stefanatos+Paspalakis:20,Stojanovic+Nauth:22,
Stojanovic+Nauth:23} 
that facilitates those technologies, have underscored the need for a fast, nearly lossless, neutral-atom transport in optically-trapped ensembles~\cite{DeChiara++:08,Torrontegui+:11,Chen+:11,Stefanatos+Li:14,
ZhangEtAl:15,Ness+:18,Lu+:20,Hickman+Saffman:20,Ding+:20,Hauck+:21,Hauck+Stojanovic:22,Norcia+:24,Hwang+:25,Morandi+:25,Cicali+:25}. As a result, coherent atom transport~\cite{Qi+:21} enabled by moving the confining optical trap -- either an optical lattice, an array of optical tweezers, or their combination -- has lately been attracting considerable attention, on the experimental- and theoretical sides alike. While the renewed experimental interest in this subject pertains largely to the issues related to increased scalability of neutral-atom QC platform~\cite{Norcia+:24}, the rekindled theoretical interest has also been motivated by the availability of powerful STA-based quantum-control protocols that can be utilized to model such transport~\cite{STA_RMP:19, STA_Early1:97, STA_Early2:99,
STA_Focus:19,ControlReview:25}.

Coherent single-atom transport in an optically-trapped system entails a final atomic state that approximates - as closely as possible - the initial state in the comoving reference frame (i.e. the rest frame of the moving optical trap), up to an irrelevant global phase. This last requirement, tantamount
to demanding minimization - ideally a complete absence - of vibrational excitations at the final point of transport, does not, however, necessarily rule out transient excitations at intermediate
times~\cite{Torrontegui+:11}. It is 
this last aspect that motivates one to design moving-trap trajectories using 
STA-based control protocols~\cite{STA_RMP:19,STA_Focus:19}. Generally speaking, such protocols typically lead to the same final states as their adiabatic counterparts, but often require significantly shorter times, thus alleviating the debilitating effects of noise and decoherence. Owing to the fact that adiabatic control-parameter variations typically render some dynamical 
properties of the system invariant, inverse-engineering schemes based on Lewis-Riesenfeld invariants 
(LRIs)~\cite{Lewis+Riesenfeld:69} constitute the most pervasive subclass of STA quantum-control protocols.

Early theoretical investigations of coherent atom transport were mainly restricted to one-dimensional 
(1D) systems~\cite{Torrontegui+:11} with purely harmonic confining potentials~\cite{Chen+:11}, a physical situation that can be treated exactly within the LRI-based STA framework. However, given that such simplifications typically do not apply in realistic
systems - where either a significant coupling between the longitudinal and transverse degrees of freedom is present, or the relevant confining potential is anharmonic~\cite{ZhangEtAl:15} - subsequent theoretical studies~\cite{Hauck+:21,Hauck+Stojanovic:22} have been devoted to more practically relevant, 
higher-dimensional optical lattices of the 
optical-conveyor-belt~\cite{Hickman+Saffman:20} 
and double-well~\cite{Stojanovic+:08,Hofer+:12} types. Atom transport in such systems is enabled by the dynamical-lattice effect, which is induced by appropriately arranged pairs of acousto-optic 
modulators (AOMs) that give rise to a frequency imbalance between two counterpropagating laser beams. These studies of atom transport in higher-dimensional optical lattices were carried out using the recently proposed eSTA method~\cite{Whitty+:20,Odelli+:23,Odelli+:24} - a predominantly analytical modification of the STA approach that allows one to account for the anharmonic character of the underlying potentials. The results obtained using this method are comparable to those corresponding to the quantum-speed limit~\cite{Hauck+Stojanovic:22}. 

In this paper, we aim to demonstrate -- using both the STA approach and eSTA-based modification -- the feasibility of 
time-efficient coherent atom transport in an adjustable family of 2D optical lattices proposed in \cite{Tarruell+:12}; this family includes square, honeycomb, dimerized, and 1D-chains lattices as its special cases. Dynamical optical lattices that belong to this 
family have already been utilized experimentally for demonstrating topological pumping~\cite{Walter+:23} 
and for realizing two-qubit quantum gates with neutral atoms~\cite{Kiefer+:25}. 

We start by proposing specific arrangements of AOMs that enable the dynamical-lattice effect in an arbitrary direction in the relevant lattice plane. We then set out to model the ensuing coherent atom transport by employing two classes of control protocols: (i) STA-type protocols, here based on inverse-engineering techniques involving LRIs, and (ii) protocols based on the eSTA method, which unlike STA-based ones systematically take into account the full optical-lattice potential. Having obtained the dynamical-lattice trajectories based on each of these two methods, we evaluate the resulting single-atom dynamics by solving the corresponding time-dependent Schr\"{o}dinger equation (TDSE) in the comoving reference frame numerically using the Fourier split-operator method (FSOM). We then quantify and compare 
the efficiency of STA- and eSTA-based atomic transport through the corresponding transport fidelities for a broad range of values of the relevant system parameters.

In the manner described above, we show that both approaches -- STA and eSTA -- allow time-efficient atom transport in the family of optical lattices under consideration. We demonstrate that while STA provides somewhat faster transport in shallow lattices (i.e. for relatively small lattice depths), eSTA becomes far superior to it for deeper lattices. Based on our numerical evaluations, we show that this superiority of the eSTA approach compared to its STA counterpart is more pronounced for shorter transport distances. 

In addition to finding both STA- and eSTA solutions to the atom-transport problem at hand, we devote due attention to analyzing their sensitivity to errors, more precisely their robustness against deviations in optical-lattice depths. The obtained results bear out the superior robustness of the eSTA protocols compared to their STA counterparts in the transport problem under consideration, which to a large extent originates from the general assumptions underlying the inner workings of the eSTA approach.

The remainder of this paper is organized as follows. We start Sec.~\ref{system} by providing the basic background of the atom-transport problem at hand, including the Hamiltonian describing such transport and the essential properties of the relevant family of optical lattices; 
we then specify concrete arrangements of AOMs that enable atom transport in the system under consideration. 
Section~\ref{ReviewSTAandESTA} is devoted to a short survey of the general STA- and eSTA-related results of relevance for the problem at hand. In the following, Sec.~\ref{MovLattTrajectory}, we describe how these general results are utilized for obtaining STA and eSTA dynamical-lattice trajectories and analyze typical examples thereof. Section~\ref{AtomDynamics} is devoted to the description of our approach for the treatment of atomic dynamics. In Sec.~\ref{ResultsDiscussion}, 
we discuss the results for the atom-transport 
fidelities obtained using either of the two approaches;
we then discuss how these results are affected by the 
presence of small lattice-depth variations, and, finally,
also demonstrate the superiority of our adopted transport schemes to some alternative approaches for modeling coherent
atom transport in optical lattices. To end with, a short summary of the paper - along with the main conclusions 
drawn - is provided in Sec.~\ref{SummaryConclusions}. In 
order not to interrupt the flow of the discussion in the 
main part of this paper, certain cumbersome mathematical derivations are relegated to Appendices~\ref{DeriveGn} 
and \ref{DeriveKn}. For similar reasons, in Appendix~\ref{ReviewFSOM} we briefly review the 
application of the FSOM for solving TDSEs, while Appendix~\ref{FSOMgroundState} contains a short 
introduction into the imaginary-time evolution (ITE) 
approach for computing ground states of quantum systems. Finally, in Appendix~\ref{DerivationSimulation} we derive
a dimensionless set of units that enables more accurate numerical simulations than the original units.

\section{System and atom transport} \label{system}
To set the stage for further discussion, we introduce 
the optical-lattice system to be considered in what 
follows and the principal objectives of our investigation.
In particular, in Sec.~\ref{SingleAtomTransport} below 
we briefly formulate the problem of atom transport and 
the characteristic spatial, temporal, and energy 
scales. We then describe the main features and 
parameters characterizing the relevant family of 2D optical lattices (Sec.~\ref{LattPotential}), followed 
by a discussion of specific configurations of AOMs 
that give rise to the dynamical-lattice effect in this 
system in different directions in the $x-y$ plane (Sec.~\ref{AOMconfig}).

\subsection{Atom transport in optical lattices} \label{SingleAtomTransport}
The problem to be discussed in what follows is that of 
transporting an atom of mass $m$ (e.g. of $^{40}$K), 
located at the time $t=0$ in one of the minima of a 2D
optical-lattice potential, 
to another one of its minima, which is situated at a distant 
location in the $x-y$ plane. In keeping with the conventional atom-transport phenomenology, we assume hereafter that the transport distance $d$ is at least an order of magnitude larger than the linear dimension (i.e. the spatial extent) $l_0$ of the atomic ground-state wave packet, that is, $d\gtrsim 10\:l_0$. The principal assumption 
inherent to atom transport in optical lattices is that an atom initially 
resides in the ground state of the potential well corresponding to a lattice-potential minimum and -- having covered the distance
$d$ at the time $t=t_f$ -- also finds itself in the ground state of the potential well at 
its destination. In other words, the atom transport 
process amounts ideally to a perfect state transfer -- from one ground state to its spatially dislocated counterpart described by the same wave function (up to a physically unobservable global phase) -- and is not being accompanied by the creation of any excitations.

It will be assumed in this work that atom transport occurs by way of the dynamical-lattice effect; in the simplest scenario this effect is enabled by the interference of two counterpropagating laser 
beams with a mutual frequency imbalance. Such an imbalance can straightforwardly be created through the use of appropriately arranged pairs of AOMs (see Sec.~\ref{AOMconfig} below).

The generic form of a time-dependent Hamiltonian describing single-atom transport in the $x-y$ plane of a 2D optical lattice with the potential $U_{\mathrm{L}}(\mathbf{r})\equiv U_{\mathrm{L}}(x,y)$ reads
\begin{equation}\label{FullSingleAtomHamiltonian}
H_{\textrm{L}}(t)=--\frac{\hbar^2}{2m}\:\left(\frac{\partial^2}{\partial x^2}+
\frac{\partial^2}{\partial y^2}\right)+ U_{\mathrm{L}}\left[\mathbf{r}-\mathbf{q}_0(t)\right]\:,
\end{equation}
where $\mathbf{q}_0(t)\equiv\left[q_0^{\mathrm{x}}(t),q_0^{\mathrm{y}}(t)\right]^\intercal$ is the 
time-dependent 2D vector describing the dynamical-lattice trajectory, i.e. the path of a potential minimum. 
The TDSE corresponding to this last Hamiltonian is typically discussed in the comoving frame. The wave function 
$\Phi(\mathbf{r},t)$ of an atom 
in this last reference frame is obtained from its lab-frame counterpart $\Psi(\mathbf{r},t)$ via a generalized Galilean transformation~\cite{GottfriedYanBOOK:03}.
More precisely,  
$\Phi(\mathbf{r},t)\equiv\hat{\mathcal{U}}(t)
\Psi(\mathbf{r},t)$, where in the Hilbert space of the system at hand the unitary operator $\hat{\mathcal{U}}(t)$ is given by (see, e.g. \cite{SLKannals:08})
\begin{equation}\label{eqUnitaryOperatorTrapFrame}
\hat{\mathcal{U}}(t) = e^{i \hat{\mathbf{p}}\cdot \mathbf{q}_0(t)/\hbar}
\:e^{-i m \hat{\mathbf{r}}\cdot\dot{\mathbf{q}}_0(t)/\hbar} \:,
\end{equation}
with $\hat{\mathbf{r}}$ and $\hat{\mathbf{p}}$  
being the single-atom (2D) position and momentum
operators, respectively.

Based on the Hamiltonian in Eq.~\eqref{FullSingleAtomHamiltonian}, in Sec.~\ref{MovLattTrajectory} below we discuss how -- using STA- and eSTA methods -- one can design trajectories that enable fast, coherent atom 
transport in the 2D optical-lattice system under consideration. In particular, both of our approaches (see Sec.~\ref{AOMconfig} below) enable 
atom transport 
along an arbitrary direction in the $x-y$ plane;  by way of example, we will mostly be interested 
in transport along the $x$-, $y$-, and diagonal directions, with the respective transport distances being denoted by $d_\mathrm{x}$, 
$d_\mathrm{y}$, and $d_\mathrm{r}$.

Before closing these preliminary considerations, it is prudent to specify the characteristic time-, length-, and energy scales
in the problem under consideration. In particular, the oscillation periods $T_\mathrm{x}\equiv 2\pi/\omega_\mathrm{x}$ and $T_\mathrm{y}\equiv 2\pi/\omega_\mathrm{y}$, corresponding to the harmonic approximation of the full optical potential 
[cf. Eq.~\eqref{eqApproxPotent} in Sec.~\ref{LattPotential} below] in the $x$- and $y$ directions, will serve as the relevant time scales of atom transport in the respective directions. Likewise, the harmonic-oscillator zero-point 
lengths $l_{\textrm{x}}\equiv\sqrt{\hbar
/(2m\omega_\mathrm{x})}$ and $l_{\textrm{y}}\equiv\sqrt{\hbar/(2m\omega_\mathrm{y})}$ will play the role of the characteristic length scales. Finally, the recoil energy 
$E_{\textrm{R}}\equiv\hbar^2 k_{\textrm{L}}^2/(2m)$, where $k_{\textrm{L}}$ is the magnitude of the relevant laser wave vector $\mathbf{k}_{\textrm{L}}$, will serve as the characteristic energy scale. 

\subsection{Adjustable family of 2D optical lattices} \label{LattPotential}
In what follows, we describe the setup resulting in an adjustable family of 2D dynamical optical lattices that will be utilized as a test bed for the study of atom transport in the remaining 
part of this work. In order to provide a more concise and 
clearer explanation of the envisioned setup, we 
hereafter restrict our discussion to the strong confining potentials in the $x-y$ plane; we neglect the minor modulations arising from the ever-present confinement of the system in the $z$ direction, which in the anticipated physical realization is close to being harmonic.

\begin{figure}[b!]
\includegraphics[width=\columnwidth]{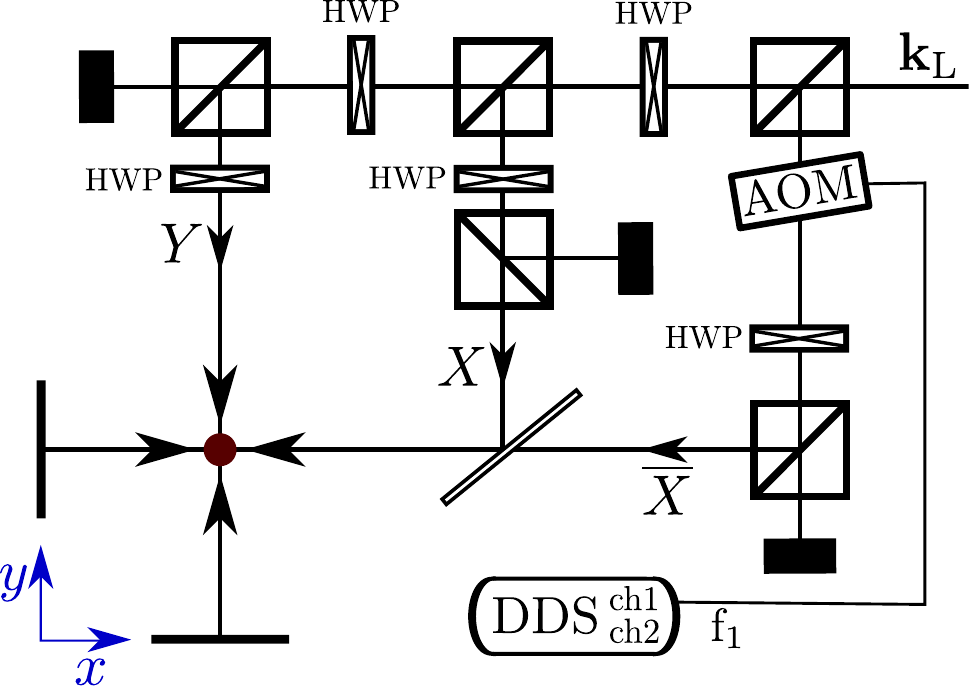}
\caption{\label{fig:Setup1}Schematic of a setup that enables
the adjustable optical-lattice formed from three folded retro-reflected beams.
The direct digital synthesizer (DDS) is used to control the frequency change induced by the acousto-optic modulator (AOM), giving rise to the phase $\theta$. The half-wave plates (HWP) can be adjusted 
in their angular orientation to set the specific intensities wanted for the three laser beams.
The red dot indicates the center of the optical-lattice potential.}
\end{figure}

The sought-after adjustable family of 2D optical lattices is constructed from an arrangement of three distinct 
retro-reflected laser beams of linear polarization and wavelength $\lambda_\mathrm{L}$~\cite{Tarruell+:12}; therefore, the entire system can be parametrized by 
a single wavenumber $k_\mathrm{L}\equiv 2\pi/\lambda_\mathrm{L}$. 
These three beams will be denoted by $\widebar{X}$, $X$, 
and $Y$ in the following (for an illustration, see 
Fig.~\ref{fig:Setup1}), 
with $I_{\widebar{X}}$, $I_X$, and $I_Y$ being the respective beam intensities. 

In the envisioned setup~\cite{Tarruell+:12}, beams $\widebar{X}$ and $X$ propagate along the same axis, while beam $Y$ is oriented along a $90^{\circ}$-rotated axis. In particular, the interference of two perpendicular beams -- namely, beams $X$ and 
$Y$ -- creates a checkerboard lattice with 
period $\lambda_\mathrm{L}/\sqrt{2}$. 
At the same time, a third beam, 
$\widebar{X}$ -- which is collinear 
with $X$, but detuned from it by a 
frequency $\delta$ -- gives rise to 
an additional standing wave with a 
period of $\lambda_\mathrm{L}/2$. 

The potential that corresponds to the envisioned adjustable family of 2D 
optical lattices 
can be derived starting from the general expression for the light intensity $I(x,y)$ in terms of the electric field $\mathbf{E}(x,y)$~\cite{ShoreBOOK:90}
\begin{equation}\label{eqIntensity}
I(x,y) = c\varepsilon_0 \:|\mathbf{E}(x,y)|^2 \:.
\end{equation}
By making use of this last expression, one finds that the intensity pattern corresponding to the checkerboard lattice is given by 
\begin{eqnarray}\label{eqIntensity1}
I_1 (x,y) =  &-&I_X \, \sin^2(k_\mathrm{L}x)  -I_Y \, \cos^2(k_\mathrm{L}y) \\
&-& 2 \gamma \, \sqrt{I_X I_Y} \sin(k_\mathrm{L}x)\cos(k_\mathrm{L}y)\cos\phi \nonumber \:,
\end{eqnarray}
where $\phi$ corresponds to the relative phase between the two arms, while $\gamma$ is 
the visibility introduced to quantify the 
influence of beam-quality in an experimental setting. The possible sources of experimental imperfections are, for instance, imperfect coherence and mode-mismatching. In what follows, we will restrict our analysis to high-contrast regimes of $\gamma=0.9$, which is readily achievable in experimental 
realizations~\cite{Tarruell+:12}.

 By once again employing the general expression for laser-beam intensity [cf. Eq.~\eqref{eqIntensity}], one finds that the laser-intensity pattern corresponding to beam $\widebar{X}$ is given by
\begin{equation}\label{eqIntensity2}
I_2 (x,y) =  - I_{\widebar{X}} \, \cos^2(k_\mathrm{L}x + \theta/2) \:,
\end{equation}
where the phase $\theta$ determines the relative position of the checkerboard 
pattern and the 1D standing wave; physically,
$\theta$ originates from a detuning between the two laser beams along the $x$ axis, 
i.e. beams $\widebar{X}$ and $X$ (for a schematic illustration, see Fig.~\ref{fig:Setup1} below). 

From Eqs.~\eqref{eqIntensity1} and \eqref{eqIntensity2} it is straightforward to obtain the expression for the total 
laser-beam intensity 
$I_{\mathrm{tot}}(x,y)= I_1
(x,y)+I_2 (x,y)$. It then immediately follows that the corresponding total optical-lattice potential 
$U_\mathrm{L}(x,y)\:\propto\: I_{\mathrm{tot}}(x,y)$, describing the envisioned family of 2D lattices, in its most general form reads
\begin{eqnarray}\label{eqOptPot}
U_\mathrm{L}(x,y) =&-& U_X \, \sin^2(k_\mathrm{L}x) -U_Y \, \cos^2(k_\mathrm{L}y) \nonumber\\
&-& U_{\widebar{X}} \, \cos^2(k_\mathrm{L}x + \theta/2) \\
&-& 2 \gamma \, \sqrt{U_X U_Y} \sin(k_\mathrm{L}x)\cos(k_\mathrm{L}y)\cos\phi \nonumber
\:,
\end{eqnarray}
where $U_X$, $U_{\widebar{X}}$, 
and $U_Y$ are the three single-beam lattice depths (proportional to the laser-beam intensities). This last optical-lattice potential is closely related to the potential utilized in the experimental realization of an adjustable family of 2D optical-lattice structures 
by Tarruell {\em et al.}~\cite{Tarruell+:12}.
For the sake of simplifying 
the analysis of coherent
atom transport in the present work,
we make the following two changes compared to \cite{Tarruell+:12}.
Firstly, we displace the $x$ coordinate by the distance $-\pi/(2k_{\mathrm{L}})$
[i.e. make the replacement $x\rightarrow x-\pi/(2k_{\mathrm{L}})$]. Secondly, we adopt a somewhat different convention for the phase $\theta$; more specifically yet, we make the replacement
$\theta\rightarrow \theta + \pi$.

\begin{figure}[t!]
\includegraphics[width=\columnwidth]{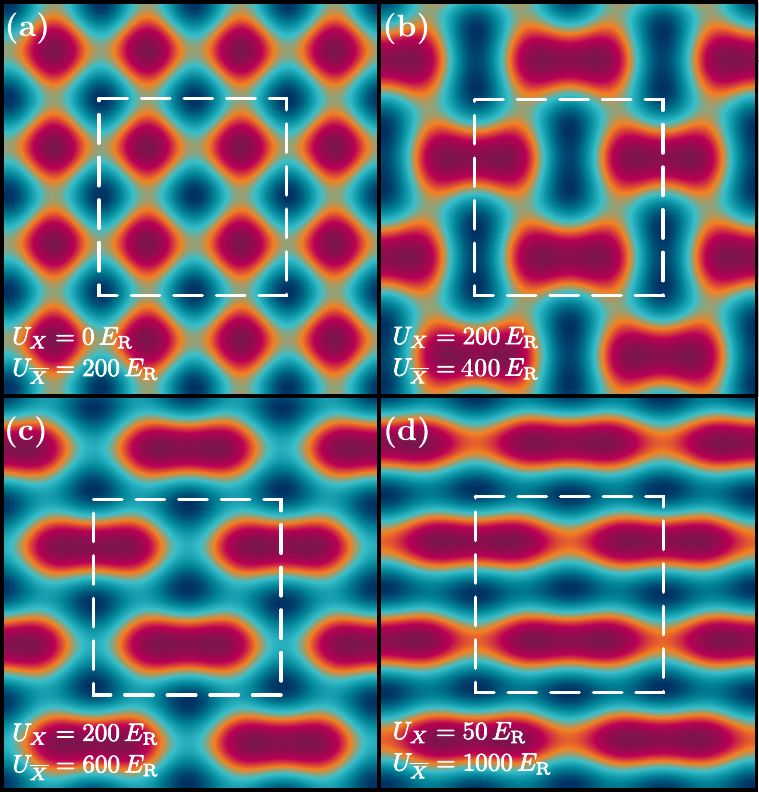}
\caption{\label{fig:DensityPlot}Density plots of the optical potential for the (a) square, (b) dimerized, (c) honeycomb, and (d) 1D-chains, optical lattices, obtainable by adjusting the different lattice depths present in the expression for $U_\mathrm{L}(x,y)$ [cf. Eq.~\eqref{eqOptPot}]. For  each lattice structure, the corresponding unit 
cell is indicated by the dashed white line. The point $(x_0, 0)$, around which the harmonic approximation of the full potential is developed, lies in the center of the unit cell.}
\end{figure}

The general optical-lattice potential in Eq.~\eqref{eqOptPot} gives rise -- by adjusting the relative magnitude of the potential depths $U_X$, $U_{\widebar{X}}$, and 
$U_Y$ -- to several different lattice configurations;
their corresponding potentials -- as well as their respective periodically-repeating units (i.e. 
unit cells) -- are visualized in 
Fig.~\ref{fig:DensityPlot}. 
Among them, the following four lattice structures can be realized: honeycomb, triangular, dimerized, and square. 
These four lattice types will be utilized in the following to test the performance of the eSTA scheme compared to its STA counterpart in modeling coherent atom transport.

Having analyzed the full, anharmonic optical-lattice potential of the system at hand [cf. Eq.~\eqref{eqOptPot}], 
it is pertinent to find at this point its harmonic counterpart, i.e. its approximate form, with terms up to quadratic order in $x$ and $y$. This harmonic potential, to be denoted by $V_{\mathrm{L}}(x,y)$, is of utmost importance for 
our further discussion of atom transport (see Secs.~\ref{HarmTrans} and \ref{MovLattTrajectory} below). 

Given that the adjustable optical-lattice potential in 
Eq.~\eqref{eqOptPot} has a minimum at the point $(x_0, y_0)$ with 
\begin{align}
x_0&= \frac{1}{k_{\mathrm{L}}} \arcsin\left(\gamma \frac{\sqrt{U_X U_Y}}{U_{\widebar{X}}-U_X}\right)\:, 
\label{eqZeroPointsPotential}\\
y_0&=0 \:, 
\end{align}
for all previously
discussed lattice configurations
(see Fig.~\ref{fig:DensityPlot}) we will approximate the full potential around their respective minima. We do so assuming that $k_\mathrm{L}|x-x_0|\ll 1$, as well as $k_\mathrm{L}|y|\ll 1$. Based on these last assumptions, it is straightforward to obtain
\begin{equation}\label{eqApproxPotent}
V_\mathrm{L}= -V_{\mathrm{d},0} + 
m a_{\mathrm{x}} x+\frac{m}{2} \left(
\omega_{\mathrm{x}}^2 x^2+\omega_{\mathrm{y}}^2 
y^2 \right) \:,
\end{equation}
where $V_{\mathrm{d},0}$ depends on 
$\theta$ and is given by
\begin{align}
V_{\mathrm{d},0}
= &U_X \sin^2(kx_0) + \frac{U_{\widebar{X}}}{2}\left[1+  \cos(2 kx_0) \cos\theta\right]\\
&+ U_Y + 2\gamma \sqrt{U_X U_Y} \sin(kx_0)\cos\phi \nonumber
\:.
\end{align}
The explicit expressions for the 
squared frequencies in Eq.~\eqref{eqApproxPotent} read 
\begin{eqnarray}
\omega_{\mathrm{x}}^2
&=& 
\label{eqHarmonicFreuency}
\frac{2 U_X k_\mathrm{L}^2}{m} 
\left(\frac{U_{\widebar{X}}}{U_X}
\cos\theta - 1
\right) \cos^2(kx_0) \:, \\
\omega_{\mathrm{y}}^2
&=&
\frac{2 U_Y k_\mathrm{L}^2}{m} \left[1+\sqrt{\frac{U_X}{U_Y}}\:\gamma \sin (k_\mathrm{L}x_0)  \cos\phi\right]
\:, \nonumber
\end{eqnarray}
while the factor $a_{\mathrm{x}}$, with dimensions of acceleration, is given by
\begin{equation}\label{eqDWOLacceleration}
a_{\mathrm{x}}= -\omega_{\mathrm{x}}^2 x_0
\:.
\end{equation}
Therefore, the approximated potential has a term linear in $x$. 
While the expansions of the conventional optical-lattice potentials do not contain such terms, these unusual terms -- linear in the spatial 
coordinate -- are also present in the case of the double-well optical lattice~\cite{Hauck+Stojanovic:22}.

Another important property of the approximated
potential $V_\mathrm{L}$ in Eq.~\eqref{eqApproxPotent}, 
of interest for our further discussion, is that -- by contrast to
the total lattice potential, which is nonseparable in the $x-y$ plane (i.e. it cannot be written as a sum of terms that depend on only on spatial coordinate)-- this approximated potential is separable; this last property greatly simplifies the analysis of atom transport governed by the approximated potential. 

\subsection{AOM configurations for atom transport} \label{AOMconfig}
\begin{figure}[b!]
\includegraphics[width=\columnwidth]{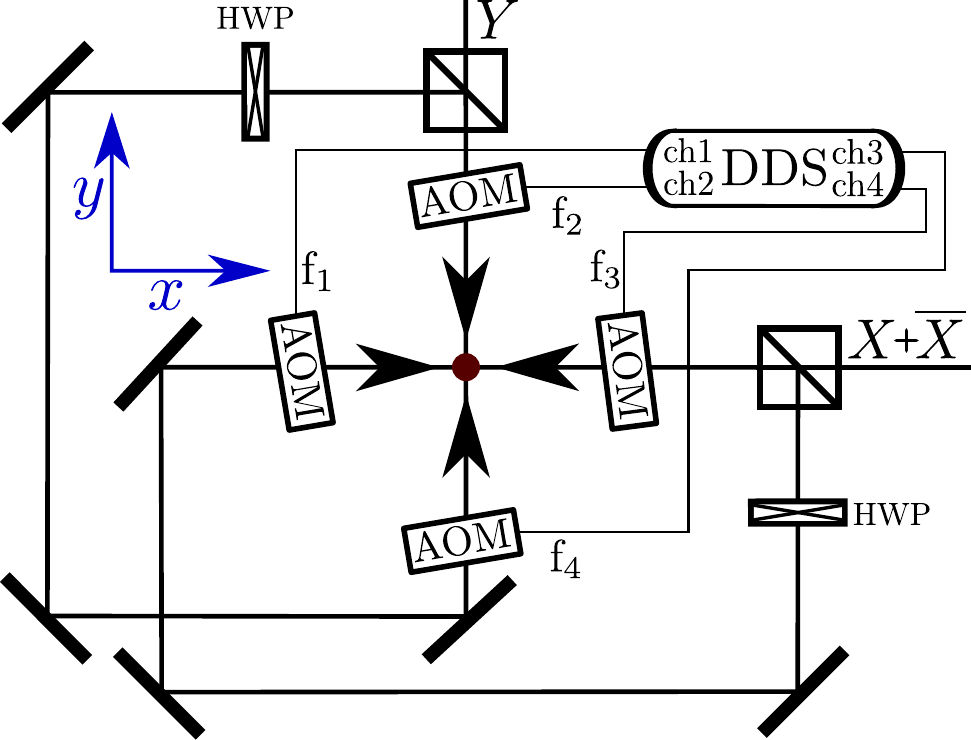}
\caption{\label{fig:Setup2}Adapted schematic of a setup that enables atom transport for the adjustable optical-lattice. The direct digital synthesizer (DDS) is used to control the frequency change induced by the acousto-optic modulator (AOM), giving rise to the dynamical-lattice velocity $v$.
The red dot indicates the center of the optical-lattice potential.}
\end{figure}

To facilitate further discussion of atom transport in the family of optical lattices under consideration, we first specify the relevant arrangements of AOMs that enable transport along $x$ and $y$ directions (for an illustration, see Figs.~\ref{fig:Setup1} 
and \ref{fig:Setup2}).

To be able to appreciate the proposed configurations of AOMs, it is helpful to first recall how the use of an
AOM gives rise to the dynamical-lattice effect. AOMs lead to a shift an incoming
frequency by applying an ultrasonic frequency onto a crystal, which results
in a density modulation
along that crystal. Accordingly, the incoming laser light encounters the 
density modulations effectively playing
the role of a grating and, consequently, leading to diffraction. Thus, the use of
the AOM leads to a frequency shift that is equal to the frequency of the ultrasonic
sound wave~\cite{Huang+:14}. In an actual optical setup, the AOMs are connected to different channels of the direct-digital-synthesizer (DDS), which allow one to control the frequencies $f_1$ and $f_2$ applied to the crystals in the AOMs. 
While one AOM turns the frequency down, 
the other one ramps it upward by the same amount; this leads to an imbalance $\Delta f=f_1-f_2=c\left(|\mathbf{k}_+|-|\mathbf{k}_-|\right)/\left(2\pi\right)$ between the counterpropagating laser beams with wavenumbers $|\mathbf{k}_+|$ and $|\mathbf{k}_-|$, respectively. This imbalance engenders a moving optical lattice with velocity 
$v =\pi \Delta f/ k_\mathrm{L}$, where $k_\mathrm{L}=|\mathbf{k}_\mathrm{L}|$ is the laser wavenumber~\cite{Hauck+Stojanovic:22}.

Having described the general procedure for incorporating AOMs into the system to enable atom transport, we now discuss how this approach can be implemented in the setup for the adjustable potential shown in Fig.~\ref{fig:Setup1}.
Currently, a single AOM is employed to tune the phase shift $\theta$ present in the optical-lattice potential of Eq. \eqref{eqOptPot}. However, it is important to clarify that this particular AOM does not participate in the shifting procedure.
To enable controlled lattice translation, each arm of the setup must first be modified so that the beams are not simply reflected by a mirror at the end of each arm. Instead, additional beam splitters and mirrors must be incorporated to separate the incoming and reflected components in each arm, allowing them to be treated independently. An illustration of the modified configuration for the relevant region is provided in Fig.~\ref{fig:Setup2}.
Once the four incoming beams are separated, individual AOMs can be placed in each path to control the frequency of the corresponding beam. In this way, a dynamical optical lattice can be generated in each arm through an induced frequency difference $\Delta f$. The resulting transport directions can then be represented as a superposition of displacements along each axis. For ideal diagonal transport, the frequency differences in both arms must be tuned identically.

\section{STA and eSTA: survey of results relevant for atom transport} \label{ReviewSTAandESTA}
To set the stage for modeling 
coherent atom transport in the family of optical lattices under consideration using the STA- and eSTA methods (see Sec.~\ref{MovLattTrajectory} below),
we first provide a short survey of these methods, with emphasis on results of relevance for atom transport. We start with a short introduction into the subclass of STA methods that employ inverse engineering based on LRIs (Sec.~\ref{BasicInvEng}) and then specialize our discussion to the problem of harmonic transport (Sec.~\ref{HarmTrans}); for the sake of simplicity, we provide an overview of these STA results using the notation that corresponds to 1D systems. Finally, we discuss the basic aspects of the eSTA method, explaining in detail how an eSTA control scheme can be obtained from its STA counterpart that is exact for a simplified Hamiltonian of the system (Sec.~\ref{eSTAbasics}).

\subsection{Invariant-based 
inverse engineering and the Lewis-Leach family 
of Hamiltonians} \label{BasicInvEng}
Inverse engineering based on LRIs - the most widely 
used subclass of STA - have by now become one of 
the preferred approach for modeling coherent atom transport. The underlying transport theory, advanced in \cite{Torrontegui+:11}, employs a particular family of quadratic-in-momentum invariants~\cite{Lewis+Riesenfeld:69}. According to that theory, the idealized harmonic-trapping case 
and that of an arbitrary (anharmonic) trapping potential entail somewhat different treatments. To be more specific, the perfect transport in the latter case requires one, in principle, to
utilize compensating forces in the comoving reference frame.

A dynamical invariant of a time-dependent Hamiltonian $H(t)$ is any operator $I(t)$ that satisfies the equation
\begin{equation}\label{eqLewisRiesenfeldInvariant}
\frac{\partial }{\partial t} I(t) + \left[H(t),I(t)\right] = 0 \:.
\end{equation}
The last equation implies that the eigenvalues $\lambda_n$ of $I(t)$ ought to be time-independent. 
If, in addition, these eigenvalues are also 
non-degenerate, the corresponding (normalized) eigenstates 
$\ket{\Phi_n(t)}$ of $I(t)$ are simply related to
the instantaneous eigenstates $\ket{\Psi_n(t)}$ 
of the Hamiltonian $H(t)$ [referred to as the transport modes of $H(t)$]. Namely, these two 
sets of states are then connected through the 
relation $\ket{\Psi_n(t)}=e^{i\theta^{\textrm{LR}}_{n}(t)}\ket{\Phi_n(t)}$, where 
\begin{equation} \label{LRphase}
\theta^{\textrm{LR}}_{n}(t)=\hbar^{-1}\int_0^t\bra{\Phi_n(t')}
\left[i\hbar\frac{\partial}{\partial t'}-H(t')\right]\ket{\Phi_n(t')} dt'
\end{equation}
is the Lewis-Riesenfeld phase~\cite{STA_RMP:19}. Consequently, the general solution of the TDSE for $H(t)$ can in this case succinctly be written as a weighted sum of transport modes, i.e.
\begin{equation}\label{eqGeneralSolutionSchroedinger}
|\Psi(t)\rangle=\sum_{n}C_n\: e^{i\theta^{\textrm{LR}}_n (t)}|\Phi_n(t)\rangle \:,
\end{equation}
where $C_n$ are constants. 

The most important family of Hamiltonians employed 
in modeling atom transport are those
of the Lewis-Leach type~\cite{Lewis+Leach:82}. These
Hamiltonians are given by
\begin{equation}\label{LewisLeachHamiltonians}
H(t)=\frac{p^2}{2m} + V(x,t)  \:,
\end{equation}
where - in the most general case - the potential $V(x,t)$ reads
\begin{equation}\label{eqPotentialLewisLeach}
V(x,t) = -F(t)x +\frac{m}{2}\:\omega^2(t) x^2 +\frac{1}{\rho^2 (t)}
U\left(\frac{x-\beta(t)}{\rho(t)}\right)  \:,
\end{equation}
where $\rho(t)$, $\beta(t)$, $\omega(t)$, and $F(t)$ are arbitrary functions of time satisfying certain auxiliary equations~\cite{Torrontegui+:11}.
In particular, the functions $\rho(t)$ 
and $\omega(t)$ ought to satisfy the nonlinear second-order ordinary differential equation 
\begin{equation}\label{AuxEq1}
\ddot{\rho}+\omega^2(t)\rho = \frac{\omega^2_0}{\rho^3} \:, 
\end{equation}
which is known as the Ermakov equation~\cite{Ermakov:1880}, where 
$\omega_0$ is a constant. At the same time, the functions $\beta(t)$, 
$\omega(t)$, and $F(t)$ 
have to obey the condition
\begin{equation}
\ddot{\beta}+\omega^2(t)\beta = \frac{F(t)}{m} \:. \label{AuxEq2}
\end{equation}
The quadratic-in-momentum dynamical invariant corresponding to the Hamiltonian in Eq.~\eqref{LewisLeachHamiltonians}, up to an immaterial additive constant, has the form~\cite{Torrontegui+:11}
\begin{eqnarray} \label{geninv}
I(t) &=& \frac{1}{2m}\big[\rho(p-m\dot{\beta})-m\dot{\rho}(x-\beta)]^2 \\
&+& \frac{m}{2}\:\omega^2_0\left(\frac{x-\beta}{\rho}\right)^2 +U\left(
\frac{x-\beta}{\rho}\right) \nonumber \:.
\end{eqnarray}
As it turns out, for a broad class of transport problems one can take $\rho(t)\equiv 1$ and $\omega(t)\equiv\omega_0$, effectively rendering Eq.~\eqref{AuxEq1} irrelevant; thus, in those cases
Eq.~\eqref{AuxEq2} remains the only auxiliary equation.

\subsection{Harmonic transport: STA treatment} \label{HarmTrans}
Having reviewed the fundamentals of invariant-based inverse engineering and its application in the special case of the Lewis-Leach-type Hamiltonians 
[cf. Eq.~\eqref{eqPotentialLewisLeach}], it is 
now pertinent to specialize our discussion to the important case of harmonic transport. In the following, we recapitulate all the relevant general results required to obtain STA transport trajectories using LRI-based inverse engineering.

In transport problems that involve 
a rigid harmonic oscillator driven by the
``transport function'' $q_0(t)$ -- corresponding 
to $F(t)=m\omega^2_0 q_0(t)$ in Eq.~\eqref{eqPotentialLewisLeach} -- the Eq.~\eqref{AuxEq2} adopts the form characteristic 
of a forced harmonic oscillator 
and $\alpha(t)$ can 
be thought of as a classical-particle trajectory
$q_c(t)$ that satisfies $\ddot{q}_c+\omega_0^2[q_c-q_0(t)]=0$.
The relevant dynamical 
invariant in the case of harmonic transport -- a special 
case of the general quadratic-in-momentum invariant 
in Eq.~\eqref{geninv} -- is given by 
\begin{equation}
I(t) = \frac{1}{2m}\big[p-m\dot{q}_c(t)]^2 \\
+ \frac{m}{2}\:\omega^2_0
\left[x-q_c(t)\right]^2 \:.
\end{equation}

In this special case of harmonic transport, the transport modes $\ket{\Psi_n(t)}=e^{i\theta^{\textrm{LR}}_{n}(t)}\ket{\Phi_n(t)}$, when written in the coordinate representation (i.e. recast in the form 
$\Psi_n(x,t)\equiv\langle x\ket{\Psi_n(t)}$, where $|x\rangle$ are the eigenstates of the position operator), are given by 
\begin{equation} \label{1DtransportModes}
\Psi_n (x,t)= e^{i\theta^{\textrm{LR}}_{n}(t)}
\Phi_{n}(x,t) \equiv e^{i\theta^{\textrm{LR}}_{n}(t)}
e^{\frac{i}{\hbar}\:m \dot{q}_c (t)x} \phi_n 
(x-q_c) \:,
\end{equation}
where $\Phi_n(x,t)\equiv\langle x\ket{\Phi_n(t)}$ and 
$\phi_n (x)$ are the harmonic eigenfunctions (i.e. solutions to the time-independent Schr\"{o}dinger equation of a 1D harmonic oscillator); the 
Lewis-Riesenfeld phase $\theta^{\textrm{LR}}_{n}(t)$
[cf. Eq.~\eqref{LRphase}] reads 
\begin{equation} \label{1DLHO_LRphase}
\theta^{\textrm{LR}}_{n}(t) =
-i\left(n+\frac{1}{2}\right)\omega_0 t 
+ \frac{m}{2}\:\int_{0}^{t}\dot{q}_{c}^{2}(t')\:dt' \:.
\end{equation}

In order to ensure that $I(t)$ and $H(t)$ are compatible at initial and final times, one sets 
$[I(t),H(t)]=0$ for $t=0$ and $t=t_f$. This translates into the following boundary 
conditions for the auxiliary function $q_c(t)$:
\begin{eqnarray} \label{BC}
q_c(0) &=& 0\:,\qquad\qquad\:\: 
q_c(t_f) = d\:, \nonumber\\
\dot{q}_c(0) &=& 
\ddot{q}_c(0) =0\:, \quad
\dot{q}_c(t_f) = 
\ddot{q}_c(t_f) =0 \:.
\end{eqnarray}
The principal idea in modeling harmonic transport is to first chose a functional form of $q_c(t)$ that satisfies these last boundary 
conditions -- typically a polynomial -- and then obtain $q_0(t)$ through inverse engineering, i.e. from the above forced-harmonic-oscillator equation. 

It is important to stress that, in addition to those in Eq.~\eqref{BC}, one can enforce further boundary conditions on $q_c(t)$ -- for example, by setting
higher-order time derivatives of 
$q_c(t)$ to zero for $t=0,\:t_f$. 
In this manner, one can construct several 
different auxiliary functions 
$q_c(t)$ for the same transport problem (e.g. 
polynomials of varying degrees) and compute 
the corresponding STA transport trajectories $q_0(t)$. This non-uniqueness 
of the resulting trajectory can be exploited
to choose a trajectory that satisfies to the largest extent possible certain additional requirements.

\subsection{Basics of the eSTA method} \label{eSTAbasics}
While STA-based control protocols have established themselves
in a variety of quantum systems~\cite{STA_RMP:19},
their modification -- which became known as eSTA -- has been proposed~\cite{Whitty+:20}. The motivation
behind this enhanced version of the STA approach, 
which is inspired in part by optimal-control methods~\cite{Whitty+:20},
is to facilitate the development of efficient control schemes for systems not directly amenable to an
STA-type treatment. The principal idea of eSTA is to approximate the original Hamiltonian of such a system 
by a simpler one that admits a straightforward 
design of an STA-based protocol. Assuming that
this last protocol is nearly optimal even when applied 
to the original Hamiltonian, its eSTA counterpart is 
constructed using a gradient expansion in the 
control-parameter space. This crucial assumption underscores the semi-heuristic character of the eSTA method, which -- in principle -- is not guaranteed 
to outperform STA~\cite{Whitty+:20}. Yet, it has already been demonstrated that eSTA is superior to STA in several classes of quantum-control problems~\cite{Hauck+Stojanovic:22,Odelli+:23}.

The initial step that the application of the eSTA method to a system described by the Hamiltonian
$H_{\textrm S}$ entails amounts to obtaining an STA solution for a ``close,'' approximated Hamiltonian $H_0$; the latter corresponds to the value $\vec{\zeta}_0$ of the vector parameter $\vec{\zeta}\in\mathbb{R}^n$~\cite{Whitty+:20}. It is assumed that a value 
$\mu_{\textrm S}$ of a scalar parameter $\mu$ can be found 
such that a series expansion 
\begin{equation} \label{eqCloseHamiltonian}
H_{\textrm S} = \sum_{k=0}^\infty \mu_{\textrm S}^k \, H^{(k)} \:,
\end{equation}
has $H_0$ as its zeroth-order term 
[i.e. $H^{(0)}\equiv H_0$].

In order to find an eSTA solution for $H_{\textrm S}$ based on a previously found STA solution for $H_0$, 
the control vector $\vec{\zeta}_{\mathrm{S}}$ corresponding to $H_{\textrm S}$ 
can be written as a sum of the
STA control vector $\vec{\zeta}_0$ and an auxiliary control vector $\vec{\alpha}$, i.e. $\vec{\zeta}_{\mathrm{S}}
=\vec{\zeta}_0 +\vec{\alpha}$. The value of 
$\vec{\alpha}$ that corresponds to the 
desired eSTA solution, will be denoted by 
$\vec{\epsilon}$ in the following.

As stated above, the principal assumption underlying the eSTA scheme is that the STA-based protocol corresponding to the Hamiltonian $H_0$ is near-optimal even when applied to the full Hamiltonian $H_{\textrm S}$~\cite{Whitty+:20}.
Another important assumption within the eSTA approach is that the deviation of the fidelity $\mathcal{F}$ from its maximal 
value is quadratic in the difference $\alpha-\epsilon$, i.e. the fidelity can be written 
in the form~\cite{Whitty+:20}
\begin{equation} \label{eqESTAFidelity}
\mathcal{F}\left(\mu_{\textrm{S}},\vec{\zeta}_0 + \alpha \frac{\vec{\nabla}\mathcal{F}(\mu_{\textrm{S}},\vec{\zeta}_0)}
{\rVert\vec{\nabla}\mathcal{F}(\mu_{\textrm{S}},\vec{\zeta}_0)\rVert}\right)\approx 1 - c\left(\alpha-\epsilon
\right)^2 \:,
\end{equation}
where $\epsilon\equiv\rVert\vec{\epsilon}\rVert$, $\alpha\equiv\rVert\vec{\alpha}\rVert$, and $c$
is a positive constant. 

The above assumptions and a Taylor expansion of the left-hand-side of Eq.~\eqref{eqESTAFidelity} around $\epsilon=\alpha$, straightforwardly lead to ~\cite{Whitty+:20}
\begin{equation}\label{eqInitialEpsilon}
\vec{\epsilon}\approx\frac{2\left[1- \mathcal{F}(\mu_{\textrm{S}},\vec{\zeta}_{\textrm{S}})\right]
\vec{\nabla}\mathcal{F}(\mu_{\textrm{S}},\vec{\zeta}_0)}{\rVert\vec{\nabla}\mathcal{F}(\mu_{\textrm{S}},
\vec{\zeta}_0)\rVert^2} \:.
\end{equation}
To second order in $\mu_{\textrm S}$ the fidelity is given by~\cite{Whitty+:20}
\begin{equation}\label{eqFidelityGn}
\mathcal{F}(\mu_{\textrm{S}},\vec{\zeta}_{\textrm{S}}) \approx
1-\frac{1}{\hbar^2}\sum_{n=1}^\infty |G_n|^2 \:,
\end{equation}
where $G_n$ is an auxiliary scalar function, given in terms of the transport modes of $H_0$ [cf. Sec.~\ref{BasicInvEng}] 
by
\begin{equation}\label{eqExpressionG}
G_n = \int_0^{t_f} dt \braket{\Psi_n(t)|
\left[H_{\textrm{S}} (\vec{\zeta}_0; t) - H_0
(\vec{\zeta}_0; t)\right]|\Psi_0(t)} \:.
\end{equation}
Likewise, up to second order in $\mu_{\textrm S}$ an analogous expression for the gradient of 
$\mathcal{F}(\mu_{\textrm{S}},\vec{\zeta}_0)$
reads~\cite{Whitty+:20}
\begin{equation}\label{eqGradientFidelityKn}
\vec{\nabla}\mathcal{F}(\mu_{\textrm{S}},\vec{\zeta}_0) \approx
-\frac{2}{\hbar^2} \sum_{n=1}^\infty \textrm{Re}
\left(G_n\:\vec{K}_n^*\right) \:,
\end{equation}
with $\vec{K}_n$ being an auxiliary vector function:
\begin{align}\label{eqExpressionK}
\vec{K}_n = \int_0^{t_f} dt \braket{\Psi_n(t)|\nabla_\zeta H_{\textrm{S}}
(\vec{\zeta};t)\big|_{\vec{\zeta}=\vec{\zeta}_0}|\Psi_0(t)} \:.
\end{align}
It should be stressed at this point that the evaluation of these two auxiliary functions ($G_n$ and $\vec{K}_n$) requires only the knowledge of the full Hamiltonian $H_{\textrm{S}}(\vec{\zeta};t)$ of the system and the known STA solutions for the idealized system governed by Hamiltonian $H_0$.

In terms of $G_n$ and $\vec{K}_n$, the eSTA correction vector $\vec{\epsilon}$ can be 
expressed  as
\begin{equation}\label{eqEpsilonDefinition}
\vec{\epsilon}= -\frac{\left(\sum_{n=1}^{N} |G_n|^2 \right) \sum_{n=1}^{N}
\textrm{Re} \left( G_n^\ast \vec{K}_n \right)}{\left\rVert \sum_{n=1}^{N} \textrm{Re}
\left( G_n^\ast \vec{K}_n \right) 
\right\rVert^2} \:,
\end{equation}
where the infinite sum over $n$ is truncated to the first $N$ terms (i.e. $N$ serves as the cut-off parameter). From this last expression, $\vec{\epsilon}$ can straightforwardly
be computed numerically provided that $G_n$ and $\vec{K}_n$ are previously calculated using Eqs.~\eqref{eqExpressionG} and \eqref{eqExpressionK}, respectively.

A general heuristic argument can be made that provides an
explanation as to why eSTA-based control protocols should be more
robust (i.e. less sensitive) to parameter deviations than their STA-based counterparts.
Namely, from the very assumption -- inherent in the construction of
the eSTA approach -- that the state fidelity behaves as a quadratic function in the vicinity of its maximum [cf. Eq.~\eqref{eqESTAFidelity}] and the fact that the fidelity 
corresponding to an eSTA scheme is  
higher than that of the
attendant STA scheme, it follows that the derivative of the 
eSTA fidelity curve should be smaller than that of its STA 
counterpart. This, in turn, signifies increased robustness of
the eSTA scheme.

\section{Dynamical-lattice trajectories}  \label{MovLattTrajectory}
In what follows, we apply the general STA- and eSTA results reviewed in Sec.~\ref{ReviewSTAandESTA} to coherent atom transport in the adjustable family of 2D optical lattices under consideration. We 
first utilize the STA approach 
to obtain the classical path of the potential minima 
in a moving lattice (see Sec.~\ref{STAtrajectory} below). We do so within the framework of a single-atom Hamiltonian with the harmonically approximated potential of Eq.~\eqref{eqApproxPotent}.
We then address the same problem using the more sophisticated eSTA approach (see Sec.~\ref{eSTAtrajectory} below),
which takes into account the full, anharmonic optical-lattice 
potential.

In the atom-transport problem under consideration, the 
$x$- and $y$ components of the (time-dependent) dynamical-lattice trajectory vector will be denoted by $q_0^{\mathrm{x}}(\vec{\zeta}_{\mathrm{x}};t)$ and $q_0^{\mathrm{y}}
(\vec{\zeta}_{\mathrm{y}};t)$. Here $\vec{\zeta}=\left(\vec{\zeta}_\mathrm{x},\vec{\zeta}_\mathrm{y}\right)^\intercal$ is the (real-valued) control vector,
where $\vec{\zeta}_{\mathrm{x}}\equiv [\zeta^{(1)}_{\mathrm{x}},\ldots,\zeta^{(6)}_{\mathrm{x}}]$
and $\vec{\zeta}_{\mathrm{y}}\equiv[\zeta^{(1)}_{\mathrm{y}},\ldots,\zeta^{(6)}_{\mathrm{y}}]$
are the reduced control vectors corresponding to the displacements in the $x$- and $y$ directions, respectively. The latter are 
assumed to satisfy the following conditions:
\begin{eqnarray}
q_0^{\mathrm{x}}(\vec{\zeta}_{\mathrm{x}};\: j t_f/7)
&=& \zeta^{(j)}_{\mathrm{x}} \quad (\:j=1,\ldots, 6\:) \:,\nonumber\\
q_0^{\mathrm{y}}(\vec{\zeta}_{\mathrm{y}};\: j t_f/7)
&=& \zeta^{(j)}_{\mathrm{y}} \:.
\end{eqnarray}

\subsection{STA dynamical-lattice trajectories} \label{STAtrajectory}
The STA dynamical-lattice trajectory vector, whose 
$x$- and $y$ components are a special case 
of $q_0^{\mathrm{x}}(\vec{\zeta}_{\mathrm{x}};t)$ 
and $q_0^{\mathrm{y}}(\vec{\zeta}_{\mathrm{y}};t)$, corresponds to the control vector $\vec{\zeta}_{0}\equiv\left(\vec{\zeta}_{\mathrm{x},0}
,\vec{\zeta}_{\mathrm{y},0}\right)^\intercal$, with 
components $\zeta^{(j)}_{\mathrm{x},0}$ and
$\zeta^{(j)}_{\mathrm{y},0}$ $(j=1,\ldots, 6)$. Thus, 
the STA trajectory is described by $q_0^{\mathrm{x}}
(\vec{\zeta}_{\mathrm{x},0};\:t)$ and 
$q_0^{\mathrm{y}}(\vec{\zeta}_{\mathrm{y},0};\:t)$.

We now proceed to apply the LRI-based transport 
theory to the approximate system Hamiltonian
\begin{equation}\label{approxHamiltonian}
H_{\textrm{L},0}(t)=-\frac{\hbar^2}{2m}\:\left(\frac{\partial^2}{\partial x^2}+
\frac{\partial^2}{\partial y^2}\right)
+V_{\mathrm{L}}(\mathbf{r}-\mathbf{q}_0(t))\:,
\end{equation}
i.e. a single-atom Hamiltonian in which $V(\mathbf{r})\equiv V(x,y)$ is given by
the simplified lattice potential $V_{\mathrm{L}}$ of Eq.~\eqref{eqApproxPotent}. This approximated potential reads
\begin{eqnarray}\label{eqDWOLSTAHamiltonian1}
& & V_\mathrm{L}\left[\mathbf{r}-\mathbf{q}_0(t)\right]= -V_\mathrm{d,0} +
m a_\mathrm{x} \big[x - q_0^{\mathrm{x}}(t)\big] \\
&+& \frac{m}{2} \big\{ \omega_\mathrm{x}^2 \big[x - q_0^{\mathrm{x}}(t)\big]^2
+\omega_\mathrm{y}^2\big[y-q_0^{\mathrm{y}}(t)\big]^2 \big\} \nonumber \:.
\end{eqnarray}

It is important to stress that -- owing to the separability of this last
potential -- displacements in the $x$- and $y$ directions can be treated independently.
Furthermore, upon restricting the above potential to one direction, i.e. making the replacement
$V_\mathrm{L} \rightarrow V^\mathrm{x}_\mathrm{L}\left(x - q_0^{\mathrm{x}}(t)\right)
\:\delta_{x,u}+ V^\mathrm{y}_\mathrm{L} \left(y - q_0^{\mathrm{y}}(t)\right)\:\delta_{y,u}
+ V_\mathrm{d,0}$ (where $u = x,\,y$), this 
potential (up to an additive constant) adopts 
the characteristic Lewis-Leach form [cf. Eqs.~\eqref{LewisLeachHamiltonians}
and \eqref{eqPotentialLewisLeach}].

In the problem under consideration, we have a Lewis-Leach-type Hamiltonian with $U=0$
and $\rho(t)\equiv 1$. As it relates to the 
remaining time-dependent parameters of the
general potential in Eq.~\eqref{eqPotentialLewisLeach}, the following
identifications can be made. In the case of atom transport in the $x$ direction, i.e. the potential
$V_\mathrm{L}^\mathrm{x}$, we set $\omega_\mathrm{x}(t) = \omega_\mathrm{x}$ and
\begin{equation}\label{eqLewisLeach1}
F_\mathrm{x}(t)= m\left[\omega^2_\mathrm{x} q_0^{\mathrm{x}}(t) - a_\mathrm{x}\right]  \:.
\end{equation}
Analogously, for atom transport along the $y$ direction, i.e. potential $V_\mathrm{L}^\mathrm{y}$,
we have $\omega_\mathrm{y}(t) = \omega_\mathrm{y}$ and
\begin{equation}\label{eqLewisLeach2}
F_\mathrm{y}(t)= m\omega^2_\mathrm{y} q_0^{\mathrm{y}}(t) \:.
\end{equation}
In keeping with the standard practice~\cite{Torrontegui+:11}, we also add to $V_\mathrm{L}^\mathrm{x}$
and $V_\mathrm{L}^\mathrm{y}$ the irrelevant time-dependent global terms $m\omega^2_\mathrm{x} q^2_{0,x}(t)$
and  $m\omega^2_\mathrm{y} q^2_{0,y}(t)$, respectively, which do not yield any force. 
Accordingly, the auxiliary equations of the forced-harmonic-oscillator type here adopt the form
\begin{align}
&\ddot{q}_\mathrm{c,x}(t) + \omega_\mathrm{x}^2
\left[ q_{\mathrm{c}}^{\mathrm{x}}(t) - q_0^{\mathrm{x}}(t) \right]=-a_\mathrm{x} \label{FHOeq1}\:,
\\
&\ddot{q}_\mathrm{c,y}(t) + \omega_\mathrm{y}^2 \left[ q_{\mathrm{c}}^{\mathrm{y}}(t)
- q_0^{\mathrm{y}}(t) \right]=0 \:, \label{FHOeq2}
\end{align}
where $\mathbf{q}_\mathrm{c}(t)\equiv\left[q_{\mathrm{c}}^{\mathrm{x}}(t),q_{\mathrm{c}}^{\mathrm{y}}(t),0\right]^\intercal$
is the corresponding 2D classical-particle trajectory.

One peculiar aspect of these last auxiliary equations 
is the presence of $a_{\mathrm{x}}$ on the
RHS of Eq.~\eqref{FHOeq1}, resulting
from the terms linear in $x$ in the approximated potentials $V_{\mathrm{L}}$ [cf. Eq.~\eqref{eqApproxPotent}]. The
presence of $a_{\mathrm{x}}$ renders the problem at hand analogous 
to transport problems involving a constant force (e.g., gravity)~\cite{Torrontegui+:11}. It should be noted that
Eq.~\eqref{FHOeq1} can be reduced to the form of Eq.~\eqref{FHOeq2} 
by a slight redefinition
of $q_{\mathrm{c}}^{\mathrm{x}}(t)$ -- namely, by introducing $q_{\mathrm{c}}^{\mathrm{x}}(t)\equiv q_{\mathrm{c}}^{\mathrm{x}}(t)
+a_{\mathrm{x}}/\omega_\mathrm{x}^2$. In this manner, the 
problem of atom transport in the $x$ direction, equally 
like the one in the $y$-direction, is reduced to a well-known 
type of atom-transport problems solvable 
via inverse engineering~\cite{Torrontegui+:11}.

\begin{figure}[b!]
\includegraphics[width=\columnwidth]{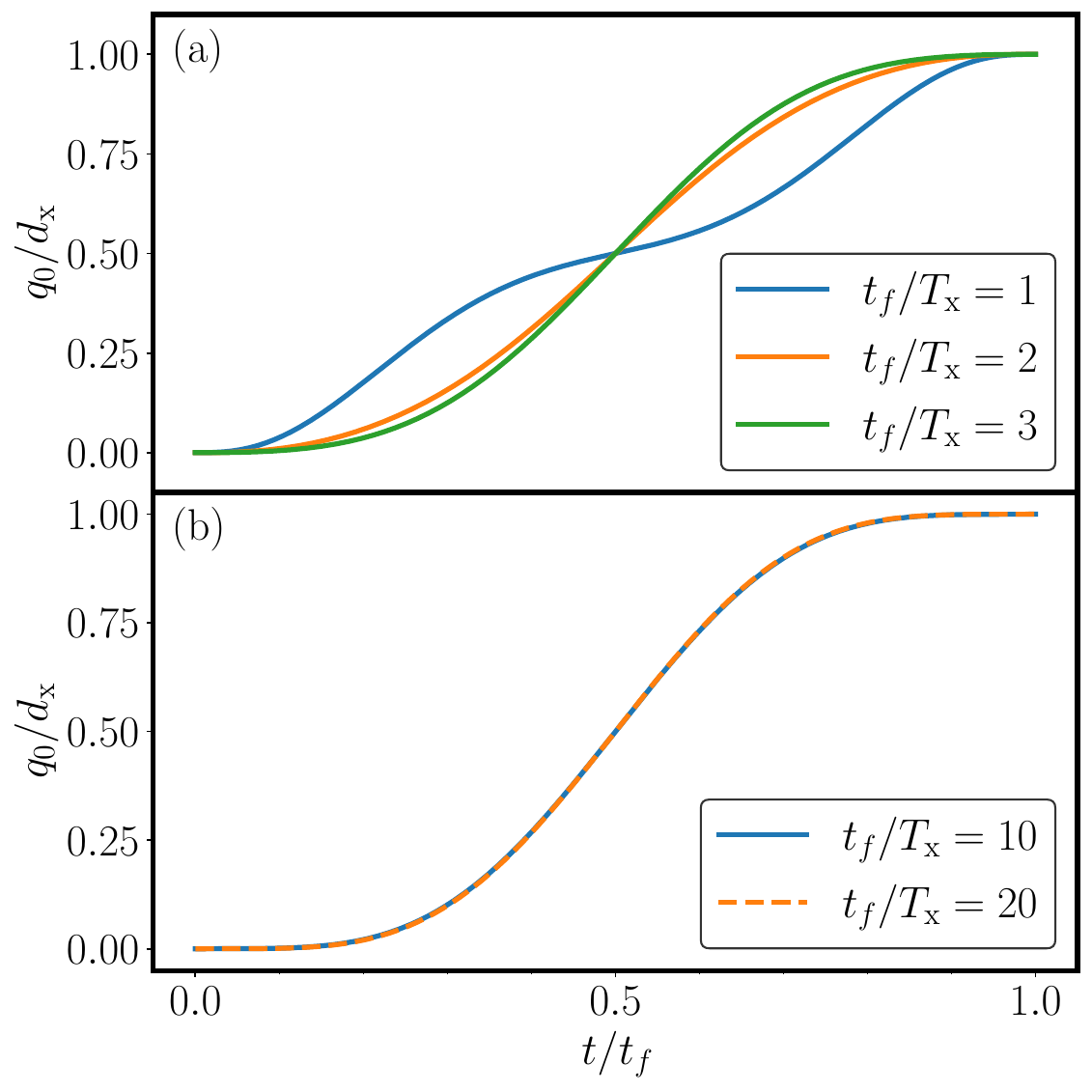}
\caption{\label{fig:STAminimum}(Color online) Path of a potential minimum as a function
of time, obtained using the STA approach, for transport times $t_f$ comparable to (a), and
an order of magnitude longer than (b) the internal timescale $T_\mathrm{x}$.
}
\end{figure}

The boundary conditions for $q_0^{\mathrm{x}}(t)$ and 
$q_0^{\mathrm{y}}(t)$ -- along
with those for for $q_{\mathrm{c}}^{\mathrm{x}}(t)$ and 
$q_{\mathrm{c}}^{\mathrm{y}}(t)$ consistent with
Eqs.~\eqref{FHOeq1} and \eqref{FHOeq2} -- motivate one to 
seek the general solution for the path of the potential 
minimum in the form of the ninth-degree polynomial~\cite{Hauck+:21}
\begin{equation}\label{STAsolution}
q_0^u(t) = d_u \sum_{n=3}^9 b_{n,u}
\left(\frac{t}{t_\mathrm{f}}
\right)^n \quad (u=x,y)\:,
\end{equation}
where $d_u$ is the transport distance. From the relevant initial/boundary conditions~\cite{Hauck+:21}, the following results for constants $b_{n,u}$ are obtained:
\begin{eqnarray}
b_{3,u} &=& 2520\:(t_\mathrm{f} \omega_u)^{-2} \quad,\quad b_{4,u}=-12600\:(t_\mathrm{f} \omega_u)^{-2}\:, \nonumber\\
b_{5,u} &=& 22680\:(t_\mathrm{f} \omega_u)^{-2} +126 \:, \nonumber \\
b_{6,u} &=& -17640\:(t_\mathrm{f} \omega_u)^{-2} -420 \:, \\
b_{7,u} &=& 5040\:(t_\mathrm{f} \omega_u)^{-2} +540 \:, \nonumber \\
b_{8,u} &=& -315 \quad,\quad b_{9,u}=70  \quad (u=x,y) \:. \nonumber
\end{eqnarray}

The obtained solution for the STA dynamical-lattice trajectory -- more precisely, the $x$ component $q_0^{\mathrm{x}}$ of the path of the potential minimum -- is shown as a function of the transport time $t_\mathrm{f}$ in Fig.~\ref{fig:STAminimum}.

\subsection{eSTA dynamical-lattice trajectories} \label{eSTAtrajectory}
Having obtained STA dynamical-lattice trajectories in the problem 
at hand (cf. Sec.~\ref{STAtrajectory}), we now proceed to find 
their STA counterparts based on the methodology reviewed in 
Sec.~\ref{eSTAbasics}. The control vectors corresponding 
to the sought-after eSTA trajectory will be denoted 
in what follows by $\vec{\zeta}_{\mathrm{x,S}}$ and $\vec{\zeta}_{\mathrm{y,S}}$; accordingly, 
this trajectory will be described by $q_0^{\mathrm{x}}(\vec{\zeta}_{\mathrm{x,S}};\:t)$ and $q_0^{\mathrm{y}}(\vec{\zeta}_{\mathrm{y,S}};\:t)$.

The eSTA paths of the potential minimum can be expressed as 
\begin{eqnarray}\label{eqControlVectorRedefined}
q_0^{\mathrm{x}}(\vec{\zeta}_{\mathrm{x,S}}; \, t) &=&
q_0^{\mathrm{x}}(\vec{\zeta}_{\mathrm{x},0}; \,t) +
f_{\mathrm{x}}(\vec{\alpha}_{\mathrm{x}}; \, t) \:, \\
q_0^{\mathrm{y}}(\vec{\zeta}_{\mathrm{y,S}}; \, t) &=&
q_0^{\mathrm{y}}(\vec{\zeta}_{\mathrm{y},0}; \,t) +
f_{\mathrm{y}}(\vec{\alpha}_{\mathrm{y}}; \, t) \:,
\end{eqnarray}
where $\vec{\alpha}:=\left(\vec{\alpha}_{\mathrm{x}},\vec{\alpha}_{\mathrm{y}}\right)^\intercal$ 
is an auxiliary control vector. These eSTA paths are assumed to obey the conditions $q_0^{\mathrm{x}}(\vec{\zeta}_{\mathrm{x,S}}; \, jt_f/7)=\zeta^{(j)}_{\mathrm{x},0}+
\alpha^{(j)}_{\mathrm{x}}$ and $q_0^{\mathrm{y}}(\vec{\zeta}_{\mathrm{x,S}}; \, jt_f/7) =
\zeta^{(j)}_{\mathrm{y},0} + \alpha^{(j)}_{\mathrm{y}}$ ($j=1,\ldots, 6$). The auxiliary functions 
$f_u(\vec{\alpha}_u;\:t)$  ($u=x,y$) ought to satisfy the following boundary conditions:
\begin{equation}
\begin{split}
& f_u(\vec{\alpha}_{u};0) = f_u(\vec{\alpha}_u ;t_f) = 0\:, \\
& f_u(\vec{\alpha}_{u}; jt_f/7 ) = \alpha^{(j)}_u \quad (\:j=1,\ldots,6\:) \:,\\
& \frac{\mrm{d^{(n)}} }{\mrm{d}t^{(n)}}f_u(\vec{\alpha}_u ;t')
|_{t'=\lbrace 0,t_f \rbrace} = 0 \quad (\:n=1,\ldots,4\:) \:.
\end{split}
\end{equation}
These conditions are chosen such that $f_{u}(\vec{\alpha}_{u};\:t)$
can be controlled through $\vec{\alpha}_{u}$, satisfying at the same time the condition of continuity. Therefore, we choose the following Ansatz based on an 
eleventh-degree polynomial:
\begin{equation}\label{eqSolutionvectorF}
f_{u}(\vec{\alpha}_{u};\:t) = \sum_{n=0}^{11}\sum_{j=1}^6
\tilde{a}^{(j)}_{u,n}\alpha^{(j)}_{u} \left( \frac{t}{t_f} \right)^n \:.
\end{equation}
The concrete values of the coefficients 
$\tilde{a}^{(j)}_{\mathrm{x},n}$ and 
$\tilde{a}^{(j)}_{\mathrm{y},n}$
in these equations can be found in Table I of \cite{Hauck+:21}.

The as yet unknown auxiliary control vector $\vec{\alpha}$, corresponding
to the desired eSTA
solution, is given by the correction vector $\vec{\epsilon}\equiv\left(
\vec{\epsilon}_{\mathrm{x}}, \vec{\epsilon}_{\mathrm{y}}\right)^\intercal$, where 
$\vec{\epsilon}_{\mathrm{x}}$ and $\vec{\epsilon}_{\mathrm{y}}$ are evaluated using Eq.~\eqref{eqEpsilonDefinition}. 
Given that in the 2D problem at hand the transport modes 
$\ket{\Psi_n(t)}$ correspond to a time-dependent 2D 
harmonic-oscillator Hamiltonian, they can be enumerated 
by two 1D quantum numbers $\{n_{\mathrm{x}},n_{\mathrm{y}}\}$. The evaluation of 
the general expressions in Eqs.~\eqref{eqExpressionG} and \eqref{eqExpressionK} in the problem under consideration 
requires the use of various properties of Hermite polynomials~\cite{ChowBOOK:00} (for details, see Appendices~\ref{DeriveGn} and \ref{DeriveKn}). 
Regarding the cut-off parameter $N$ in Eq.~\eqref{eqEpsilonDefinition}, we adopt the value $N=2$, even if our numerical evaluations show that it is sufficient to take $N=1$.

\section{Calculation of atomic dynamics} \label{AtomDynamics}
In what follows, we briefly present our chosen approach for evaluating the single-atom dynamics corresponding to 
STA- and eSTA-based dynamical-lattice trajectories [cf. Sec.~\ref{MovLattTrajectory}].
This approach is based on the use of the FSOM for numerically solving the relevant TDSE (for 
a short introduction into this method, 
see Appendix~\ref{ReviewFSOM}).

The crucial prerequisite for computing single-atom dynamics in the problem at hand amounts to evaluating the ground-state wave-function of the relevant optical-lattice potential. Within the general framework of the atom-transport problem (cf. Sec.~\ref{SingleAtomTransport}), that wave function represents the initial ($t=0$) atomic wave packet, i.e. the initial condition for single-atom transport. This wave function can be obtained to requisite accuracy using the plane-wave-expansion method (for details, see, e.g., \cite{Stojanovic+:08}), which entails Fourier expansions of the underlying lattice potential and the periodic part of single-particle Bloch wave functions in terms of reciprocal-lattice vectors. Yet, the same can be accomplished by employing the FSOM in conjunction with the ITE approach (for details, see Appendix~\ref{FSOMgroundState}). Given that in this study we employ the FSOM for the dynamics part, this last scheme appears as the more convenient choice.

Given the character of the problem that we are concerned with, it is convenient to switch from the lab frame to the comoving frame
via a generalized Galilean transformation
[cf. Eq.~\eqref{eqUnitaryOperatorTrapFrame}]. In this manner, the lab-frame single-atom 
TDSE
\begin{equation}
i\hbar\:\frac{\partial}{\partial t}\Psi(\mathbf{r},t)=
\left[\frac{\hat{\mathbf{p}}^2}{2m}+U_\mathrm{L}(\mathbf{r}-
\mathbf{q}_0(t))\right]\Psi(\mathbf{r},t) \:,
\end{equation}
is transformed into its counterpart in the comoving frame 
\begin{eqnarray}\label{eqSchroedingerequationTrapFrame}
i\hbar\:\frac{\partial}{\partial t}\Phi(\mathbf{r},t) &=&
\Big[\frac{\hat{\mathbf{p}}^2}{2m}+\frac{m}{2}\:\dot{\mathbf{q}}_{0}^2(t)
+ m \ddot{\mathbf{q}}_0\cdot \mathbf{q}_0 \nonumber \\
&+& m \mathbf{r}\cdot\ddot{\mathbf{q}}_0(t)+U_\mathrm{L}(\mathbf{r})
\Big]\Phi(\mathbf{r},t) \:.
\end{eqnarray}
It is important to underscore
that the terms $m\dot{\mathbf{q}}_{0}^2(t)/2$ and $m\ddot{\mathbf{q}}_0(t)
\cdot\mathbf{q}_0(t)$ in Eq.~\eqref{eqSchroedingerequationTrapFrame} 
merely lead to time-dependent global
phase factors; the latter are immaterial for the atom-transport problem under consideration and can, accordingly, hereafter be safely neglected. Consequently, the total potential that a transported atom experiences in the comoving frame is given by 
\begin{equation}
W_\mathrm{L}(\mathbf{r},t)\equiv U_\mathrm{L}
(\mathbf{r})+m\mathbf{r}\cdot
\ddot{\mathbf{q}}_0(t)\:.
\end{equation}

In the problem at hand, we employ the FSOM for calculating the final atomic state, i.e. the state that results from a center-of-mass displacement of the atomic wave packet by a certain distance. The presence of an explicit time dependence in the potential 
$W_\mathrm{L}(\mathbf{r},t)$ implies that the exact 
time-evolution operator of the system 
is given by the most general expression involving a time-ordered product. However, by employing
the symmetric (or Strang) splitting~\cite{Speth+:2013} approach to time evolution [cf. Eq.~\eqref{SymmetricTrotter} in Appendix~\eqref{ReviewFSOM}]
for $\hat{A}=U_\mathrm{L}(\hat{\mathbf{r}})$ and 
$\hat{B}=\hat{\mathbf{p}}^2/(2m)+m\hat{\mathbf{r}}\cdot\ddot{\mathbf{q}}_0(t)$ the relevant time-stepping scheme in the problem under consideration [cf. Eq.~\eqref{eqFSOMBCH} in Appendix~\eqref{ReviewFSOM}] reduces to
\begin{eqnarray}\label{eqFSOM_DWOL}
\Psi(\vec{r},t&+&\delta t) = \exp\left[-\frac{i}{\hbar}\:U_\mathrm{L}(\vec{r})
\frac{\delta t}{2}\right]T_{\mathbf{r},\mathbf{p}}(\delta t)\nonumber\\
&\times& \exp\left[-\frac{i}{\hbar}\:U_\mathrm{L}(\vec{r})\frac{\delta t}{2}\right]
\Psi(\vec{r},t)+ \mathcal{O}(\delta t^3)  \:,
\end{eqnarray}
where $\hat{T}_{\mathbf{r},\mathbf{p}}(\delta t)$ stands for the operator 
\begin{equation}\label{TrpDef}
\hat{T}_{\mathbf{r},\mathbf{p}}(\delta t)\equiv\exp\left[-\frac{i}{\hbar} 
\left(\frac{\hat{\mathbf{p}}^2}{2m}\:\delta t+m\hat{\mathbf{r}}\cdot\int_t
^{t+\delta t}\ddot{\mathbf{q}}_0(t')\mathrm{d}t'\right)\right] \:.
\end{equation}
By making use of the Baker-Campbell-Hausdorff formula and its special case known as the Weyl identity~\cite{GilmoreBOOK:12},
$\hat{T}_{\mathbf{r},\mathbf{p}}(\delta t)$ 
can be recast as (for a detailed derivation, see Appendix C in \cite{Hauck+Stojanovic:22})
\begin{eqnarray} \label{TrpFinal}
\hat{T}_{\mathbf{r},\mathbf{p}}(\delta t) &=& \exp\left[-\frac{i}{\hbar} 
\frac{\hat{\mathbf{p}}^2}{2m}\:\delta t \right] \,
\exp\left[-\frac{i}{2\hbar}\:\hat{\mathbf{p}}\cdot\delta\dot{\mathbf{q}}_0(t)\:\delta t\right] \, \nonumber\\
&\times& \exp\left[-\frac{i}{\hbar}\:m\hat{\mathbf{r}}\cdot\delta\dot{\mathbf{q}}_0(t)\right] \:,
\end{eqnarray}
where $\delta\dot{\mathbf{q}}_0(t)\equiv \dot{\mathbf{q}}_0(t+\delta t)-\dot{\mathbf{q}}_0(t)$.
It is worthwhile to mention that two additional multiplicative terms -- that do not depend on the momentum- 
and coordinate operators, hence leading to irrelevant time-dependent global phase 
factors -- have been dropped
on the right-hand-side (RHS) of the 
last equation, analogously to the 
dropped terms in Eq.~\eqref{eqSchroedingerequationTrapFrame}
above.

With the aid of an analog of Eq.~\eqref{eqFourierRepresentation} and by performing a Fourier transformation of 
the $\hat{\mathbf{p}}$-dependent terms in \eqref{TrpFinal}, switching also to the coordinate representation 
of the momentum operator [i.e. $\hat{\mathbf{p}}\rightarrow (\hbar/i)\nabla$], we arrive at the 
FSOM-based (second-order) time-stepping scheme for the atom-transport problem at hand:
\begin{eqnarray}\label{eqFSOMFinal}
\Phi(\mathbf{r},t&+&\delta t)= \exp\left[-\frac{i}{\hbar}\:U_\mathrm{L}(\mathbf{r})
\frac{\delta t}{2}\right]\mathrm{F}^{-1}\Bigg[e^{-i\frac{\hbar k^2}{2m}
\delta t} \nonumber \\
&\times& \exp\left[-\frac{i}{2}\:\mathbf{k}\cdot\delta\dot{\mathbf{q}}_0\delta t\right]
\mathrm{F}\Bigg\{\exp\left[-\frac{i}{\hbar}\:m\mathbf{r}\cdot\delta\dot{\mathbf{q}}_0\right] \nonumber\\
&\times& \exp\left[-\frac{i}{\hbar}\:U_\mathrm{L} (\mathbf{r}) \frac{\delta t}{2}\right]
\Phi(\mathbf{r},t)\Bigg\}\Bigg] + \mathcal{O}(\delta t^3)  \:.
\end{eqnarray}

In order to alleviate the computational burden in the problem at hand, we exploit the discrete translational symmetry of the system and restrict ourselves to the displacement of one unit cell of the original optical lattice. 
The appropriate choice of 
this unit cell has to be consistent
with the transport direction. For instance, in order to evaluate transport in the $x$- and $y$ directions
we make use of a rectangular unit cell (indicated by white dashed lines in Fig.~\ref{fig:DensityPlot}).
At the same time, describing transport in the diagonal direction necessitates the use of a $45$-degree rotated
square-shaped unit cell (black dashed lines in Fig.~\ref{fig:DensityPlot}).

Yet another problem-specific feature that allows one to further reduce the level of computational difficulty  
in the problem at hand is closely related to its description in the comoving frame. 
Namely, the form of the relevant TDSE [cf. Eq.~\eqref{eqSchroedingerequationTrapFrame}] immediately implies that there is no need to compute the potential term after each time step, but only the correction term $m\mathbf{r}\cdot\ddot{\mathbf{q}}_0$ [cf. Eq.~\eqref{TrpDef}] that 
depends on the acceleration $\ddot{\mathbf{q}}_0$ of the potential minimum.

To carry out the spatial Fourier transformation 
using the FFT algorithm (cf. Appendix~\ref{ReviewFSOM})~\cite{NRcBook}, 
we make 
use of a discrete two-dimensional $N_{\mathrm{x}}\times N_{\mathrm{y}}$ grid with $N_{\mathrm{x}}=N_{\mathrm{y}}=300$ points in the computational window whose dimensions (in units of $l_{\mathrm{x}}$ and $l_{\mathrm{y}}$) are $80$ and $80$, respectively. 
As regards the time-domain propagation 
of the TDSE in the comoving frame [cf. Eq.~\eqref{eqSchroedingerequationTrapFrame}], we carry it out using the adaptive 
approach~\cite{NRcBook} with a maximal relative error of $10^{-4}$, which could -- strictly speaking -- require additional 
(smaller) substeps in time. Yet, in all our runs, it proved sufficient to take $N_t$ between $20$ and $100$.

\section{Results and discussion} \label{ResultsDiscussion}
In what follows, we quantify and discuss the efficiency of STA- and eSTA-based coherent single-atom transport 
(cf. Secs.~\ref{STAtrajectory} and \ref{eSTAtrajectory} above, respectively)
in various optical lattices belonging to the lattice family under consideration.
The main figure of merit used for this purpose is the transport fidelity, defined as
$\mathcal{F}(t_f)
=|\braket{\Psi_{\textrm{target}}|\Psi(t_f)}|^2$. Its dependence on the transport time $t_f$ is determined by the overlap of the target state $|\Psi_{\textrm{target}}\rangle$ (the ground state of the displaced optical-lattice potential, computed using the FSOM combined with the ITE approach; 
see Appendix~\ref{FSOMgroundState}) and 
the actual final (time-evolved) atomic state 
$|\Psi(t_f)\rangle$ (evaluated using FSOM-based 
numerical propagation of the TDSE in the comoving 
frame; see Sec.~\ref{AtomDynamics} above and Appendix~\ref{ReviewFSOM}).

In Sec.~\ref{TranspFidelities} 
below, we present our results for 
atom-transport fidelities. In Sec.~\ref{Robustness} we then 
discuss how these results are 
affected in the presence of small variations in lattice depths. Finally,
in Sec.~\ref{CompareAdHoc} we demonstrate the superiority of 
our adopted transport schemes compared to approaches based on ad-hoc velocity profiles.

\subsection{Atom-transport fidelities} \label{TranspFidelities}
\begin{figure}[t!]
\includegraphics[width=\columnwidth]{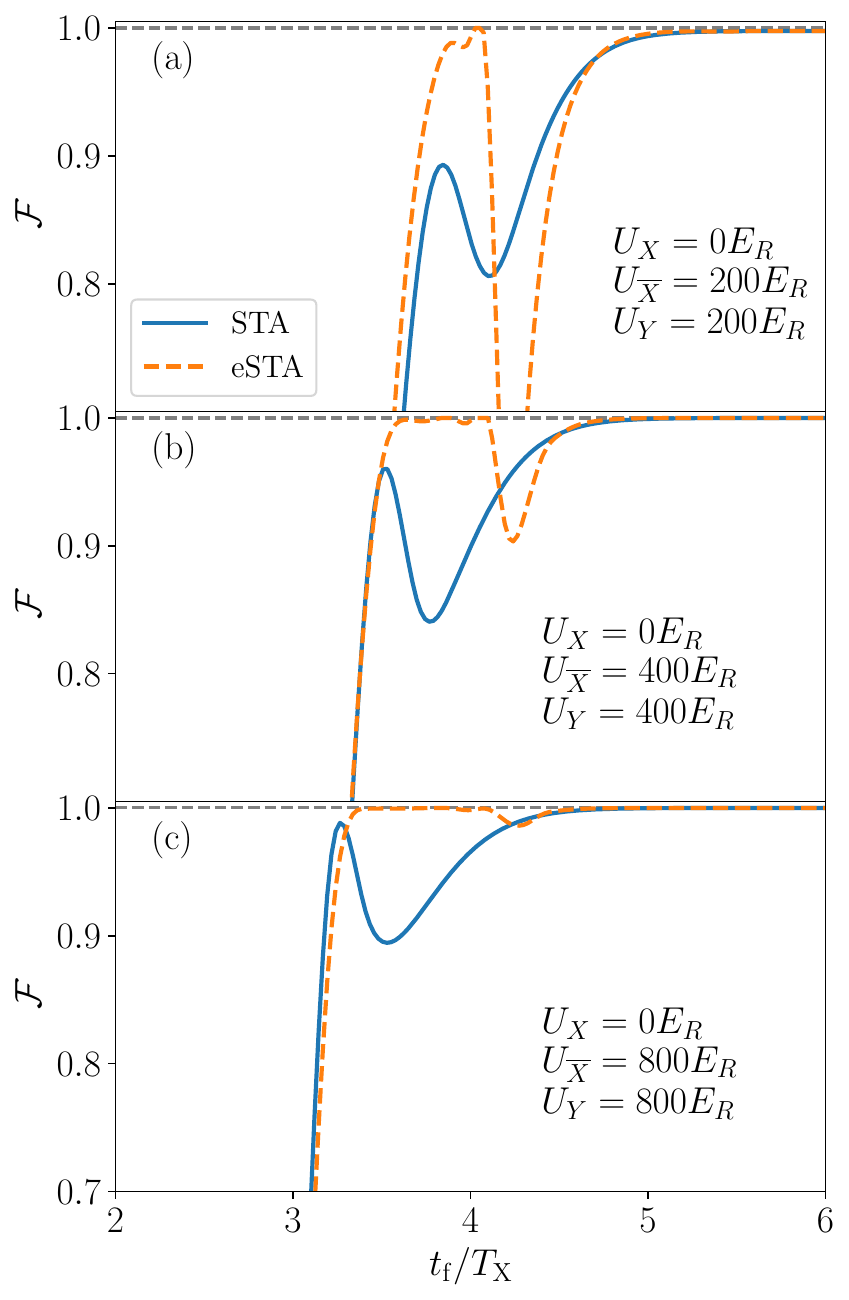}
\caption{\label{fig:FidelityResultsSquareX}(Color online) Dependence of the STA- and eSTA atom-transport fidelity on the transport time $t_\mathrm{f}$ for a transport distances of $d_{\mathrm{x}}=100\,l_\mathrm{x}$ and fixed potential depths. The ratio of the lattice depths corresponds to that of a square lattice
[cf. Fig.~\ref{fig:DensityPlot}(a)].
Panels (a)-(c) show results for potential depths of $(U_X, U_{\overline{X}}, U_Y) = (0, 200, 200) E_R$, $(0, 400, 400) E_R$, and $(0, 800, 800) E_R$, respectively.
}
\end{figure}

\begin{figure}[t!]
\includegraphics[width=\columnwidth]{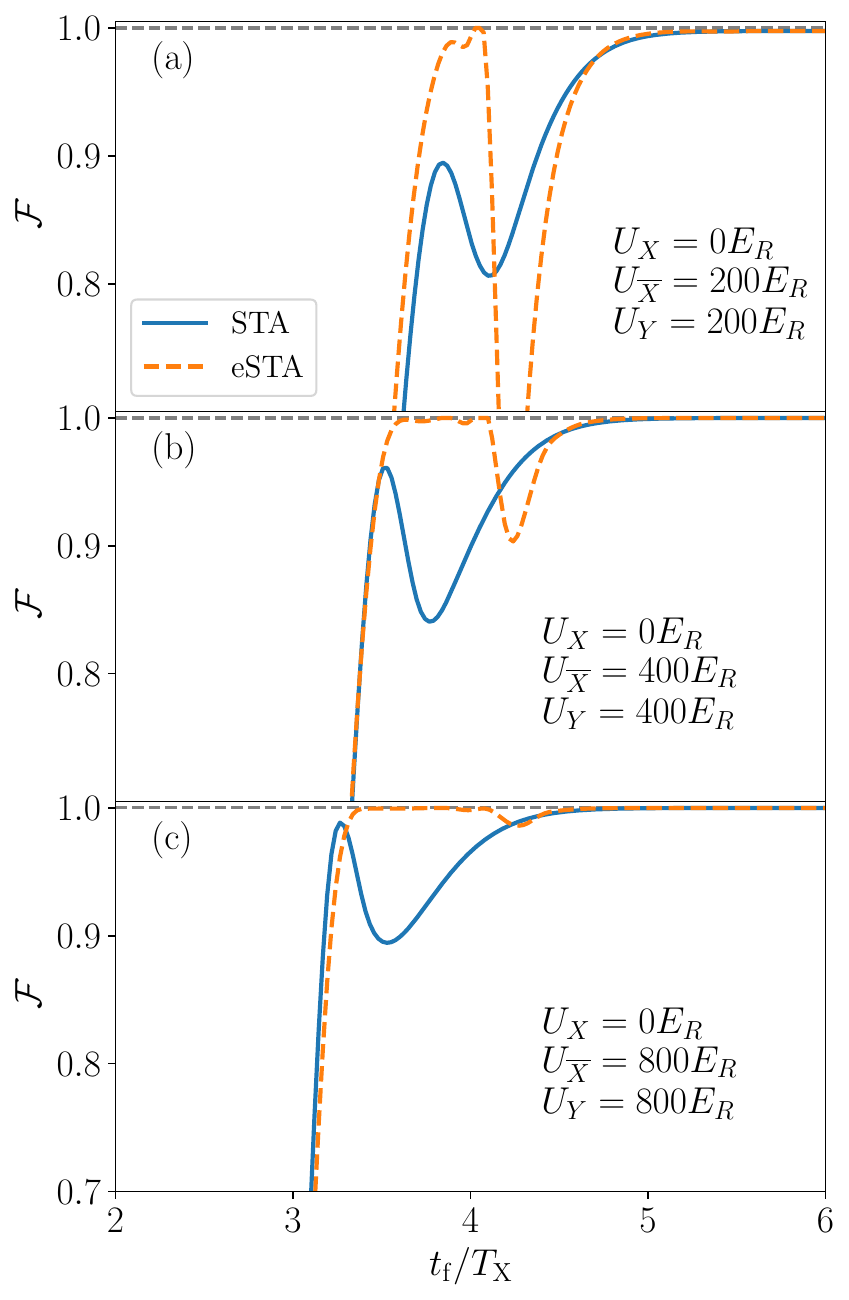}
\caption{\label{fig:FidelityResultsSquareY}(Color online) Dependence of the STA- and eSTA atom-transport fidelity on the transport time $t_\mathrm{f}$ for a transport distances of $d_{\mathrm{y}}=100\,l_\mathrm{y}$ and fixed potential depths. The ratio of the lattice depths corresponds to that of a square lattice
[cf. Fig.~\ref{fig:DensityPlot}(a)].
Panels (a)-(c) show results for potential depths of $(U_X, U_{\overline{X}}, U_Y) = (0, 200, 200) E_R$, $(0, 400, 400) E_R$, and $(0, 800, 800) E_R$, respectively.
}
\end{figure}

\begin{figure}[t!]
\includegraphics[width=\columnwidth]{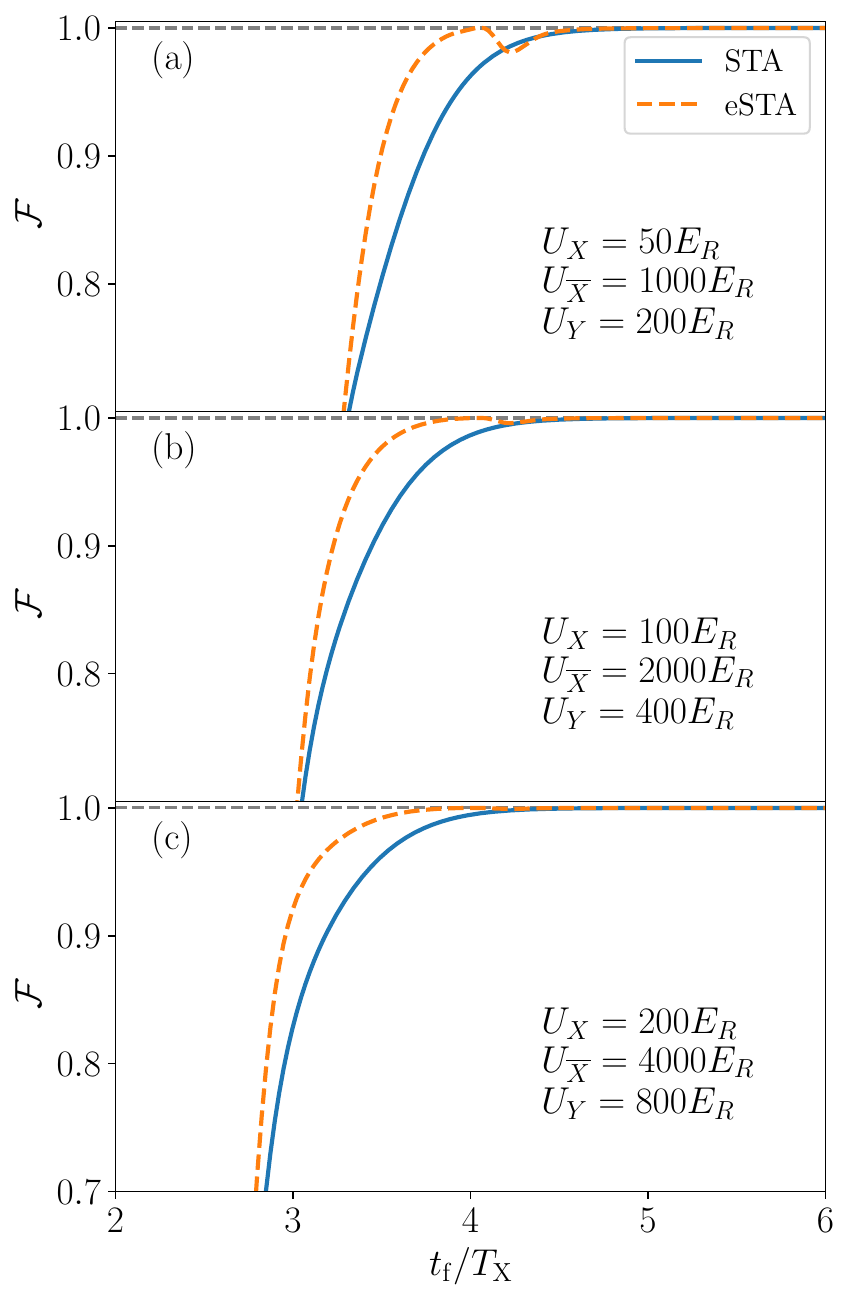}
\caption{\label{fig:FidelityResultsLineX}(Color online) Dependence of the STA- and eSTA atom-transport fidelity on the transport time $t_\mathrm{f}$ for a transport distances of $d_{\mathrm{x}}=100\,l_\mathrm{x}$ and fixed potential depths. The ratio of the lattice depths corresponds to that of the 1D-chains lattice [cf. Fig.~\ref{fig:DensityPlot}(d)]. Panels (a)-(c) show results for potential depths of $(U_X, U_{\overline{X}}, U_Y) = (50, 1000, 200) E_R$, $(100, 2000, 400) E_R$, and $(200, 4000, 800) E_R$, respectively.
}
\end{figure}

\begin{figure}[t!]
\includegraphics[width=\columnwidth]{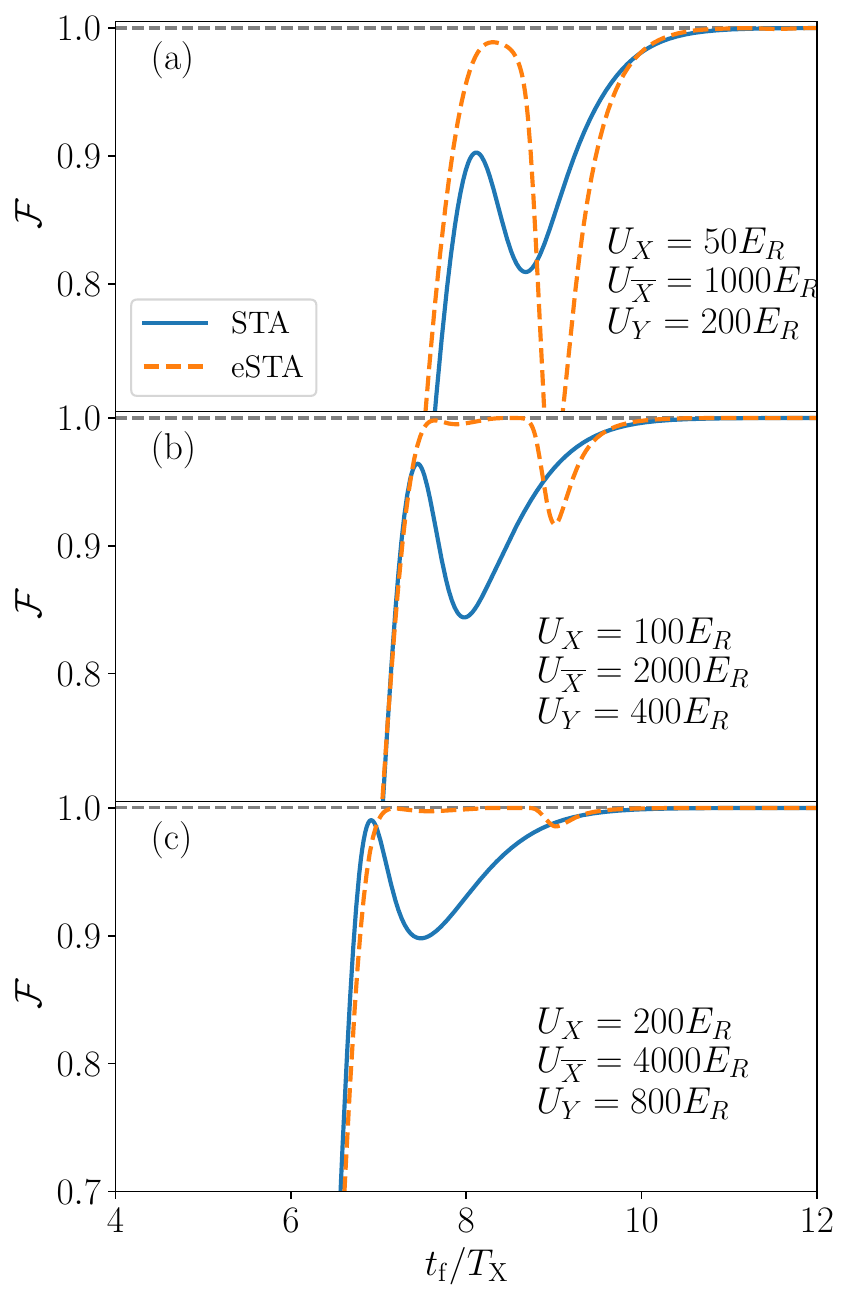}
\caption{\label{fig:FidelityResultsLineY}(Color online) Dependence of the STA- and eSTA atom-transport fidelity on the transport time $t_\mathrm{f}$ for a transport distances of $d_{\mathrm{y}}=100\,l_\mathrm{y}$ and fixed potential depths. The ratio of the lattice depths corresponds to that of the 1D-chains lattice [cf. Fig.~\ref{fig:DensityPlot}(d)]. Panels (a)-(c) show results for potential depths of $(U_X, U_{\overline{X}}, U_Y) = (50, 1000, 200) E_R$, $(100, 2000, 400) E_R$, and $(200, 4000, 800) E_R$, respectively.
}
\end{figure}

\begin{figure}[t!]
\includegraphics[width=\columnwidth]{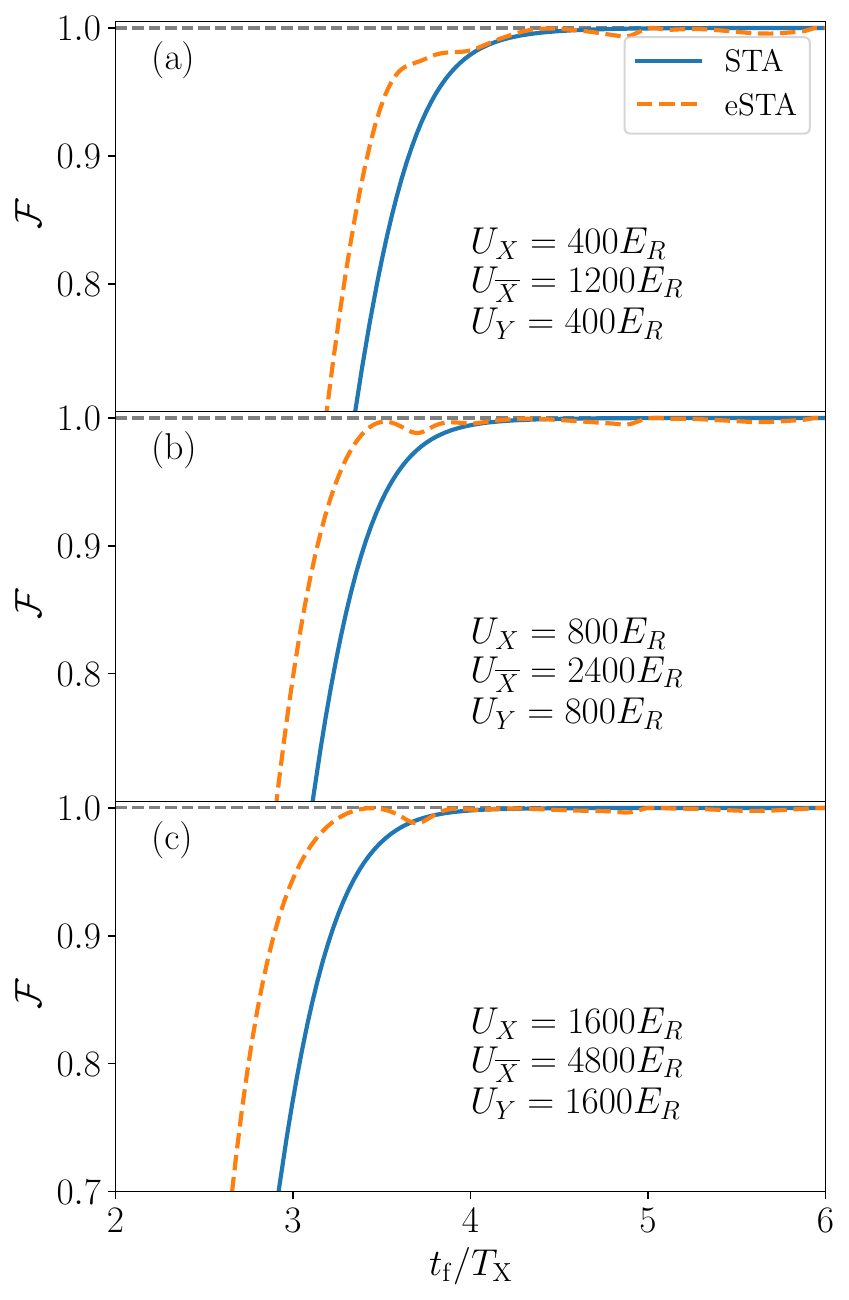}
\caption{\label{fig:FidelityResultsHoneycombX}(Color online) Dependence of the STA- and eSTA atom-transport fidelity on the transport time $t_\mathrm{f}$ for a transport distances of $d_{\mathrm{x}}=50\,l_\mathrm{x}$ and fixed potential depths. The ratio of the lattice depths corresponds to that of a honeycomb lattice [cf. Fig.~\ref{fig:DensityPlot}(c)]. Panels (a)-(c) show results for potential depths of $(U_X, U_{\overline{X}}, U_Y) = (400, 1200, 400) E_R$, $(800, 2400, 800) E_R$, and $(1600, 4800, 1600) E_R$, respectively.
}
\end{figure}

\begin{figure}[t!]
\includegraphics[width=\columnwidth]{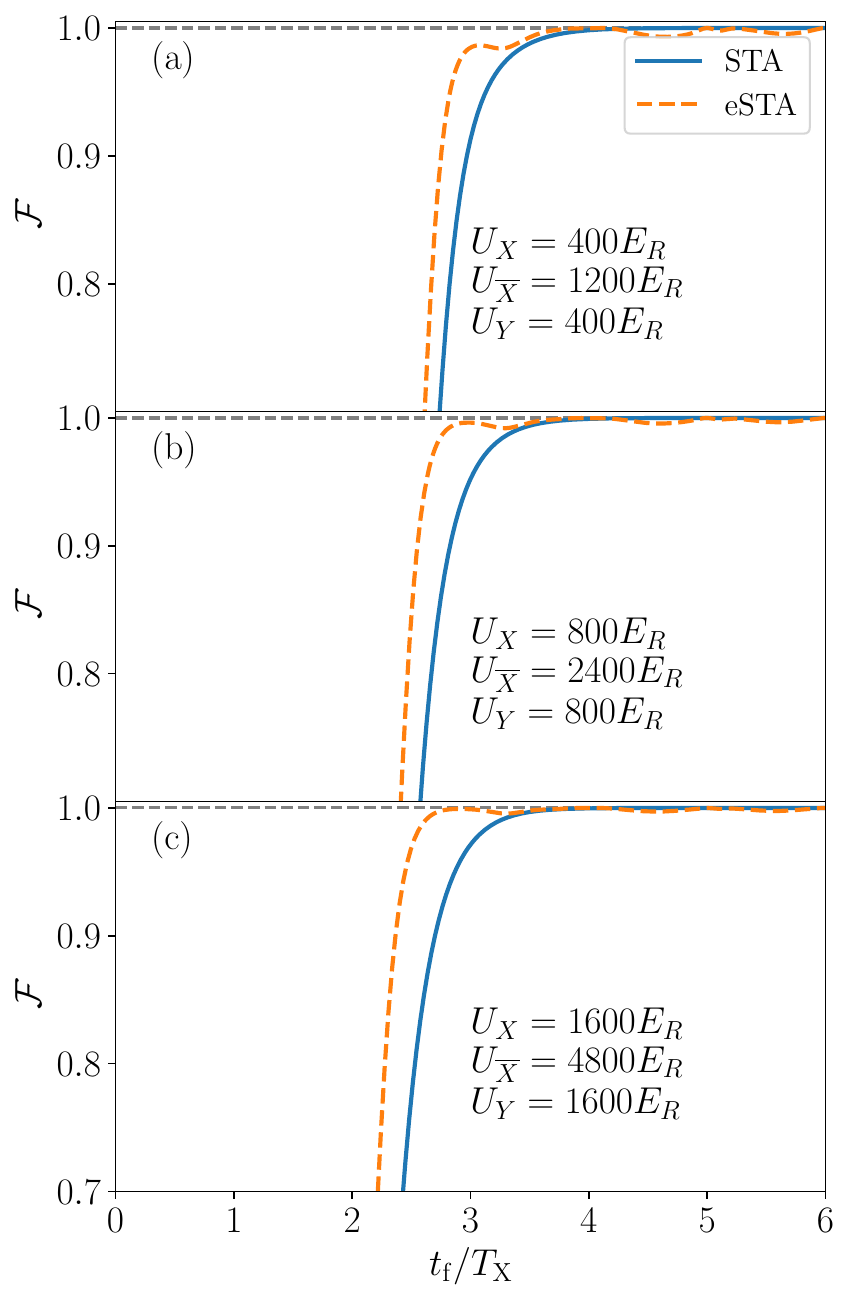}
\caption{\label{fig:FidelityResultsHoneycombY}(Color online) Dependence of the STA-and eSTA atom-transport fidelity on the transport time $t_\mathrm{f}$ for a transport distances of $d_{\mathrm{y}}=50\,l_\mathrm{y}$ and fixed potential depths. The ratio of the lattice depths corresponds to that of a honeycomb lattice [cf. Fig.~\ref{fig:DensityPlot}(c)].
Panels (a)-(c) show results for potential depths of $(U_X, U_{\overline{X}}, U_Y) = (400, 1200, 400) E_R$, $(800, 2400, 800) E_R$, and $(1600, 4800, 1600) E_R$, respectively.
}
\end{figure}

\begin{figure}[t!]
\includegraphics[width=\columnwidth]{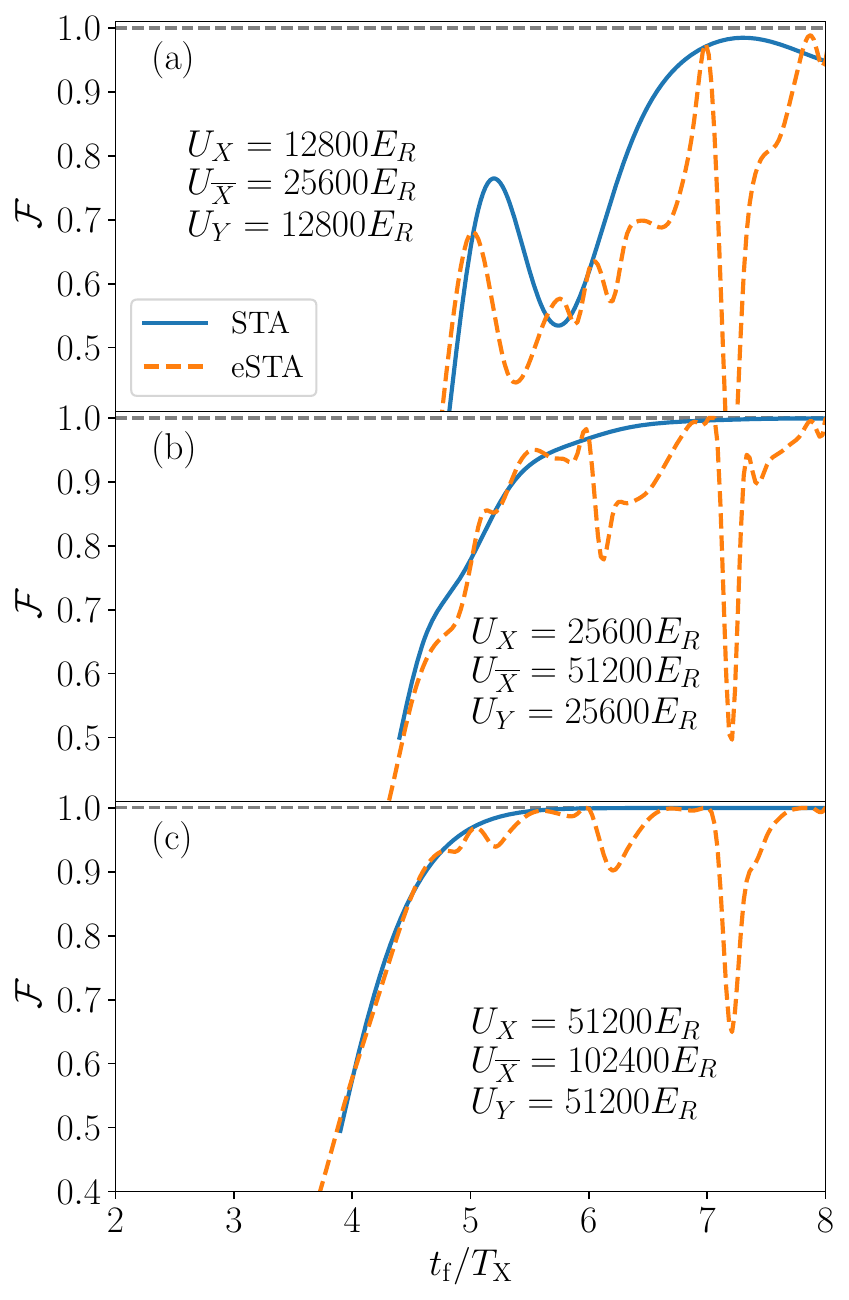}
\caption{\label{fig:FidelityResultsDimerX}(Color online) Dependence of the STA- and eSTA atom-transport fidelity on the transport time $t_\mathrm{f}$ for a transport distances of $d_{\mathrm{x}}=50\,l_\mathrm{x}$ and fixed potential depths. The ratio of the lattice depths corresponds to that of a dimer lattice
[cf. Fig.~\ref{fig:DensityPlot}(b)].
Panels (a)-(c) show results for potential depths of $(U_X, U_{\overline{X}}, U_Y) = (12800, 25600, 12800) E_R$, $(25600, 51200, 25600) E_R$, and $(51200, 102400, 51200) E_R$, respectively.
}
\end{figure}

\begin{figure}[t!]
\includegraphics[width=\columnwidth]{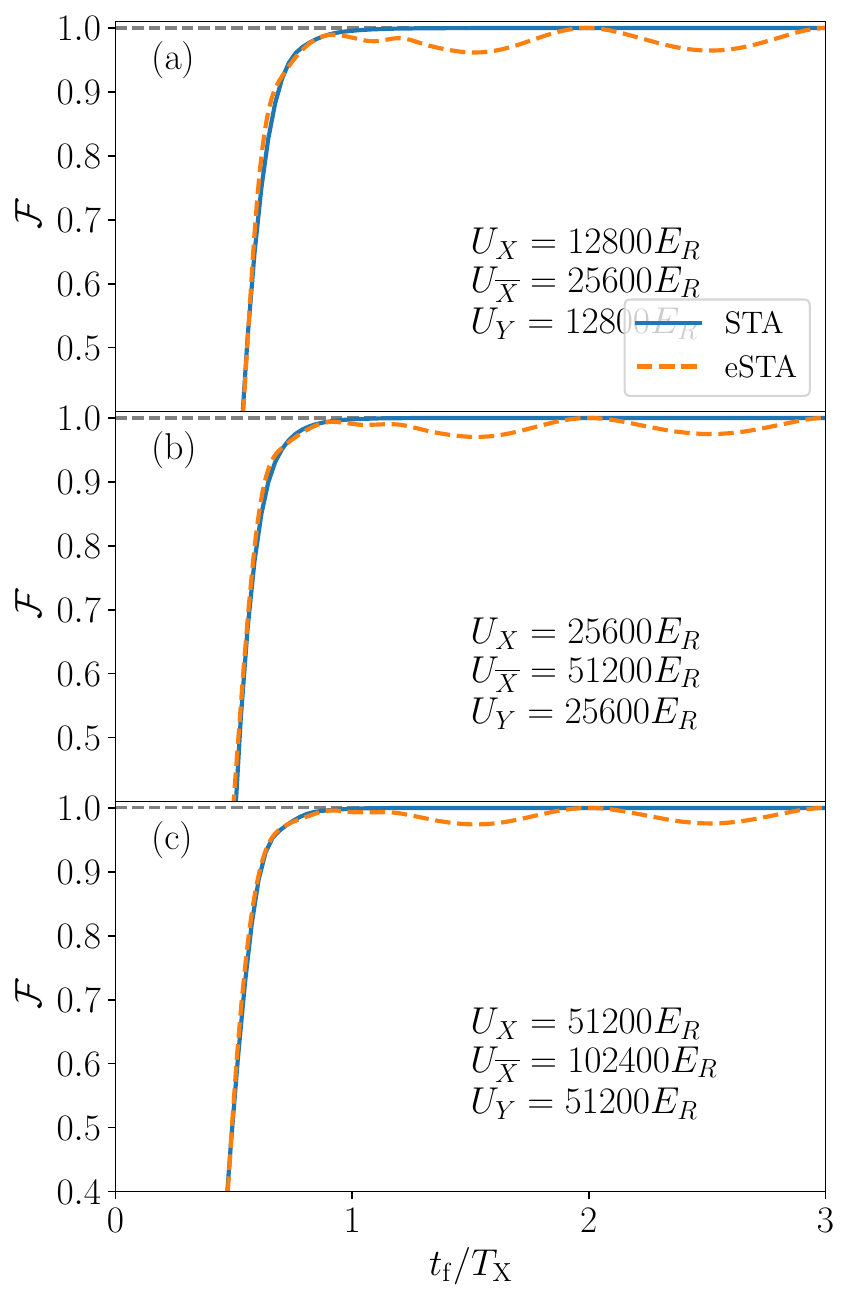}
\caption{\label{fig:FidelityResultsDimerY}(Color online) Dependence of the STA- and eSTA atom-transport fidelity on the transport time $t_\mathrm{f}$ for a transport distances of $d_{\mathrm{y}}=50\,l_\mathrm{y}$ and fixed potential depths. The ratio of the lattice depths corresponds to that of a dimer lattice 
[cf. Fig.~\ref{fig:DensityPlot}(b)].
Panels (a)-(c) show results for potential depths of $(U_X, U_{\overline{X}}, U_Y) = (12800, 25600, 12800) E_R$, $(25600, 51200, 25600) E_R$, and $(51200, 102400, 51200) E_R$, respectively.
}
\end{figure}

The obtained results for the transport fidelity in the optical lattices under consideration are depicted
in Figs.~\ref{fig:FidelityResultsSquareX} -- \ref{fig:FidelityResultsDimerY}. One common denominator of all these results, irrespective of the chosen values of the system parameters (such as the lattice depth, transport distance, beam waists, etc.), is the collapse of transport -- manifested by a precipitous decrease in the fidelity -- for very short transport times
$t_{\mathrm{f}}$. This characteristic breakdown of transport 
originates from the fact that for atomic accelerations above a certain maximal value $|a_{\textrm{max}}|$, which is proportional to the lattice depth~\cite{Hauck+:21}, the 
atom being transported cannot be 
confined by the optical-lattice potential. 
This happens as a consequence of the disappearance of the 
minima of the ``tilted'' lattice potential experienced by the atom in its comoving (non-inertial) reference frame; compared to the potential in the lab frame, the 
comoving-frame potential  
has an additional term 
linear in the transport coordinate~\cite{Hauck+:21}. This characteristic behavior for short transport times is a generic 
property of coherent atom transport in optical lattices, i.e. it is not special for the concrete family of lattices being considered here.

The collapse of transport actually takes place when the maximal acceleration of an atom during the transport process, to be hereafter denoted by $|\tilde{a}_{\textrm{max}}|$, exceeds $|a_{\textrm{max}}|$. 
Needless to say, the actual value of $|\tilde{a}_{\textrm{max}}|$ depends upon the chosen dynamical-lattice trajectory (i.e. the path of the 
potential minimum);
however, it holds quite generally that $|\tilde{a}_{\textrm{max}}| \propto\:t_{\mathrm{f}}^{-2}$. 
What follows from this last result is that for deeper lattices, which then also implies a larger value of 
$|a_{\textrm{max}}|$ because the latter is proportional to the lattice depth, the largest atomic acceleration $|\tilde{a}_{\textrm{max}}|$ reached exceeds 
$|a_{\textrm{max}}|$ for sufficiently short transport 
times $t_{\mathrm{f}}$. In other words, for deeper lattices the collapse of transport (i.e., the precipitous decrease 
of the transport fidelity) takes place for shorter times $t_{\mathrm{f}}$. Our numerical results are consistent with this general argument, as can be verified for gradually increasing lattice depths in all figures independent on the specific transport direction or lattice configuration.

In the opposite limit of very long transport times $t_\mathrm{f}$, the STA results converge to the well-known adiabatic regime. In this limit, essentially perfect atom transport ($\mathcal{F}\approx 1$) can be achieved for $t_\mathrm{f}\gg\omega_\mathrm{u}^{-1}$ ($u=x,y$). However, this behavior is only partially recovered in the eSTA approach. In cases where the potential minimum does not reside at the origin---corresponding to the presence of an additional linear term in the harmonic approximation, cf. Eq.~\eqref{eqDWOLacceleration}---the adiabatic limit is reached only for sufficiently deep lattice potentials, as discussed for the honeycomb and dimer potentials at the end of this section.

The transport fidelity corresponding to the square lattice for displacements along the $x$-direction is depicted in Fig.~\ref{fig:FidelityResultsSquareX}. While the eSTA fidelity curve in this case exhibits a pronounced dip for shallow lattice potentials in panel~(a), it nevertheless achieves a significantly improved fidelity compared to its STA counterpart for transport times $t_\mathrm{f}\approx 3.8\,T_\mathrm{X}$. This improvement becomes increasingly pronounced for deeper potentials, culminating in an almost uniformly high fidelity close to unity in panel~(c). The remaining minor deviations are a shallow dip for $t_\mathrm{f}\approx 4.2\,T_\mathrm{X}$ and a slightly earlier onset of fidelity degradation for $t_\mathrm{f}\approx 3.2\,T_\mathrm{X}$. For even deeper lattices, these features are expected to vanish completely, yielding a consistently improved transport performance through the eSTA scheme for the square lattice. Owing to the high degree of symmetry of this lattice configuration [cf. Fig.~\ref{fig:DensityPlot}(a)], an analogous behavior is observed for displacements along the $y$-direction, as illustrated in Fig.~\ref{fig:FidelityResultsSquareY}.

For displacements along the $y$-direction in the 1D-chains lattice the corresponding STA- and 
eSTA fidelity curves are shown in Fig.~\ref{fig:FidelityResultsLineY}. What can be inferred from these results is that their qualitative behavior closely follows that of the square lattice. The most notable differences are the minimal achievable 
high-fidelity transport times, which are shifted to $t_\mathrm{f}\approx (6.5-7.5)\,T_\mathrm{X}$, and the increased lattice depth 
of $U_\mathrm{\widebar{X}}$, reaching values up to $4000\,E_R$. While this depth is substantially larger than in the square-lattice case, it remains within reach of state-of-the-art experimental optical-lattice setups.

By contrast to atom transport in 
the $y$ direction, transport along 
the $x$-direction of the 1D-chains lattice [for an illustration,
see Fig.~\ref{fig:FidelityResultsLineX}, exhibits a qualitatively different behavior. Rather than displaying a sharp dip followed by an abrupt loss of fidelity, the STA transport fidelity is characterized by a more gradual degradation as the transport time $t_\mathrm{f}$ decreases. This can be attributed to the more slowly varying potential depth along the 
$x$-direction. Remarkably, the eSTA 
transport fidelity already provides a noticeable improvement for relatively shallow lattice depths in panel~(a), albeit with a minor dip for $t_\mathrm{f}\approx 4.3\,T_\mathrm{X}$. For increasing lattice depths, eSTA consistently outperforms STA across the entire range of transport times, where even the previously described dip in fidelity is not visible anymore. 
It is important to notice that the sharp fidelity drop for the eSTA case observed in other geometries for small $t_\mathrm{f}$ is absent here; instead, eSTA transport fidelity exhibits a smoother decay as well, closely mirroring the gradual behavior of its STA counterpart.

The STA results for transport along the $x$-direction of the honeycomb lattice, shown in Fig.~\ref{fig:FidelityResultsHoneycombX}, similarly display a gradual reduction in transport fidelity for shorter transport times and do not exhibit a pronounced dip prior to the final drop. In addition, small oscillations are observed for $t_\mathrm{f}\approx 4.8\,T_\mathrm{X}$ and $t_\mathrm{f}\approx 5.5\,T_\mathrm{X}$ for all considered lattice depths, panels~(a)--(c). These oscillations take place at times that are independent of the lattice depth, while their amplitude decreases for deeper potentials. These features can be attributed to the non-vanishing linear term in the harmonic approximation, Eq.~\eqref{eqDWOLacceleration}, which induces minor instabilities in the eSTA-based transport scheme, particularly for shallower lattices.

A similar discussion applies to transport along the $y$-direction of the honeycomb lattice, illustrated 
by the transport-fidelity curves in Fig.~\ref{fig:FidelityResultsHoneycombY}. Although the fidelity drop at short transport times occurs more abruptly and at slightly smaller values of $t_\mathrm{f}$ [more precisely, for $t_\mathrm{f}
\approx (2.2-2.8)\: T_\mathrm{X}$], the eSTA approach nevertheless outperforms the STA scheme in this regime.

Finally, transport in the dimer lattice along both the $x$- and $y$-directions, respectively illustrated by the fidelity curves in Figs.~\ref{fig:FidelityResultsDimerX} and \ref{fig:FidelityResultsDimerY}, remains unstable even for extremely deep lattice potentials. This behavior can be partially attributed to the enhanced contribution of the linear term in the harmonic approximation, or, equivalently, to the large values of $x_0$ for this potential configuration [cf. Eq.~\eqref{eqZeroPointsPotential}]. 
These effects give rise to oscillations that are responsible for the fidelity dips observed for $t_\mathrm{f}\approx 6\,T_\mathrm{X}$ and $t_\mathrm{f}\approx 7.1\,T_\mathrm{X}$. In addition, 
the exceptionally shallow potential barrier separating the two dimer sites plays an important role. As illustrated in Fig.~\ref{fig:DensityPlot}(b), this barrier is extremely shallow and increases only marginally with increasing lattice depth. Although eSTA exhibits some improvement for deeper lattices, it does not surpass the STA performance in this case. We expect that eSTA would eventually outperform STA for even deeper potentials; however, computational limitations prevent us from accessing this regime even further.

These effects are most pronounced for eSTA-based transport along the $x$-direction, while being noticeably improved -- but not entirely suppressed -- for displacements along $y$, as demonstrated by the eSTA results in Fig.~\ref{fig:FidelityResultsDimerY}. 
Thus, although transport in the $y$-direction exhibits enhanced stability compared to the $x$-direction, residual instabilities remain and do not vanish completely. By contrast, STA-based transport does not exhibit instabilities for the dimer lattice in either the $x$- or $y$-direction.

It should be emphasized that the eSTA results for the dimer lattice require unrealistically deep potentials that are far beyond current experimental capabilities. Consequently, it serves primarily as an academic test case to elucidate the challenging regimes of the eSTA protocol. For this reason, the dimer lattice will not be considered further in the subsequent sections, which focus on experimentally relevant configurations.

\begin{figure*}[t!]
\includegraphics[width=\textwidth]{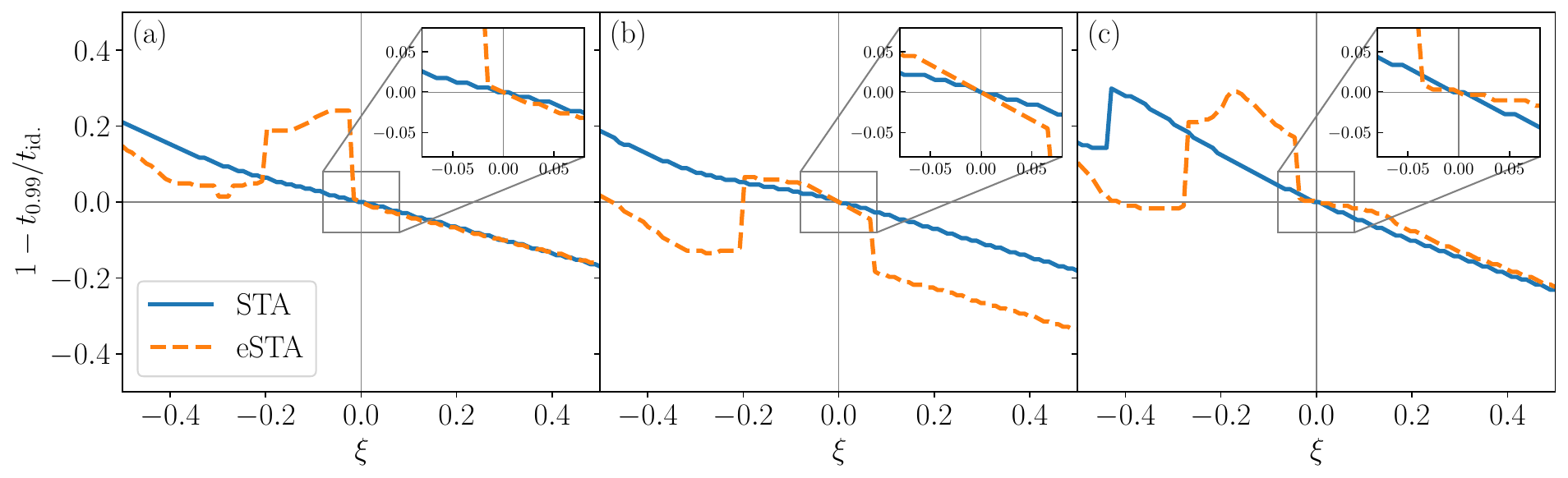}
\caption{\label{fig:StabResults} 
(Color online) Relative stability to reach an atom-transport fidelity of $0.99$, defined as $1 - t_{\mathrm{real}}/t_{\mathrm{id.}}$, 
as a function of the instability parameter $\xi$. 
Here, $t_{0.99}$ is the transport time that first drops below $0.99$ for a given $\xi$ while $t_{\mathrm{id.}}$ is the reference time $t_{0.99}$ for $\xi = 0$. 
Curves are shown for both the STA and eSTA schemes, illustrating how the transport time is affected by lattice instabilities. 
The lattice depth ratios correspond to (a) the square lattice, with $(U_X, U_{\overline{X}}, U_Y) = (0, 800, 800)\,E_R$; (b) the 1D-chains lattice, with $(U_X, U_{\overline{X}}, U_Y) = (100, 2000, 400)\,E_R$; (c) the honeycomb lattice $(U_X, U_{\overline{X}}, U_Y) = (1600, 4800, 1600)\,E_R$. The transport distance is $d_{\mathrm{x}} = 100\,l_\mathrm{x}$ for the square and 1D-chains lattice, and  $d_{\mathrm{x}} = 50\,l_\mathrm{x}$
for the honeycomb lattice.}
\end{figure*}

\subsection{Robustness against deviations in lattice depths} \label{Robustness}
To account for possible imperfections in envisioned experimental implementations of our
proposed transport schemes, here we address 
the sensitivity of those schemes to deviations in lattice depths. It is pertinent to assume that in an actual implementation of our STA- and eSTA transport schemes the nominal potential depths $U_j$ ($j \in \{X, \overline{X}, Y\}$) take slightly changed values $\widetilde{U}_j = U_j \left(1 +\xi \right)$, where $\xi$ denotes the instability parameter. 

To simplify our analysis, we keep 
$\widetilde{U}_j$ fixed at some $\xi$-independent values and make use of those values as the actual potential depths in our evaluation of STA- and eSTA transport fidelities; we previously design STA- and eSTA dynamical-lattice 
trajectories (recall Sec.~\ref{MovLattTrajectory})
with $U_j(\xi)\equiv\widetilde{U}_j/(1+\xi)$ playing the role of potential depths for each of the considered values of $\xi$. The advantage of this approach is that by fixing the potential depth the maximal acceleration supported by the potential remains the same for all data points (i.e. for different values of $\xi$), which facilitates a clearer assessment of sensitivity to deviations. While in the previous section a direct, one-to-one comparison between STA and eSTA schemes was central, here the focus lies on the individual stability properties of each scheme under parameter variations.

The results of our analysis are illustrated in Fig.~\ref{fig:StabResults}, where the relative deviation
\(
1 - t_{0.99} / t_{\mathrm{id.}}
\)
is plotted as a function of the instability parameter \(\xi\). Here, \(t_{0.99}\) denotes the transport time at which the fidelity first drops below \(0.99\), while the reference time \(t_{\mathrm{id.}}\) corresponds to the respective scheme (STA or eSTA) evaluated at \(\xi = 0\). By construction, all curves pass through the origin.

A notable trend in the STA results, visible across all three panels, is that increasing the instability parameter leads to a reduced transport time \(t_{0.99}\), resulting in an approximately constant negative slope. This behavior can be attributed to the fact that decreasing \(\xi\) effectively reduces the lattice depths and, consequently, the frequencies of the harmonic approximations underlying the STA construction. At shorter transport times \(t_\mathrm{f}\), where higher excited states contribute more strongly, a lower effective frequency provides a better approximation of the potential far away from its minimum. In this sense, reducing the frequency can be interpreted as a heuristic improvement that captures the overall potential shape more faithfully than a harmonic approximation that is better suited around the origin.

An exception to the aforementioned trend is 
found in the honeycomb-lattice case shown in panel (c), where a sudden increase in \(t_{0.99}\) occurs around \(\xi \approx -0.4\). This behavior likely originates from the fact that the harmonic frequencies are becoming too large to reliably represent the global structure of the underlying potential.

Overall, the eSTA results follow a similar qualitative trend as the STA scheme, which is expected given that eSTA is constructed as a potential tailored version build on STA. Nevertheless, additional features emerge. For the square lattice in panel (a), a pronounced reduction of \(t_{0.99}\) is observed near \(\xi = 0\), extending into the regime of negative instability parameters till around $-0.2$. In this interval, the eSTA protocol consistently outperforms the reference time \(t_{\mathrm{id.}}\), before reverting to the general trend. Since the eSTA scheme is more strongly tailored to the reference potential, and is thus expected to be more sensitive to deviations, this region of enhanced stability does not extend over the entirety of the probed regime.

Similar behavior is observed for the remaining lattice geometries. Notably, for the 1D-chains lattice shown in panel (b), the point \(\xi = 0\) itself lies within the region of improved stability. This indicates that incorporating small deviations from the reference potential in the construction of the control protocol can enhance robustness, although the extent and location of this improvement depend sensitively on the specific lattice geometry.

\subsection{Comparison with atom transport based on 
ad-hoc velocity profiles} \label{CompareAdHoc}

\begin{figure*}[t!]
\includegraphics[width=\textwidth]{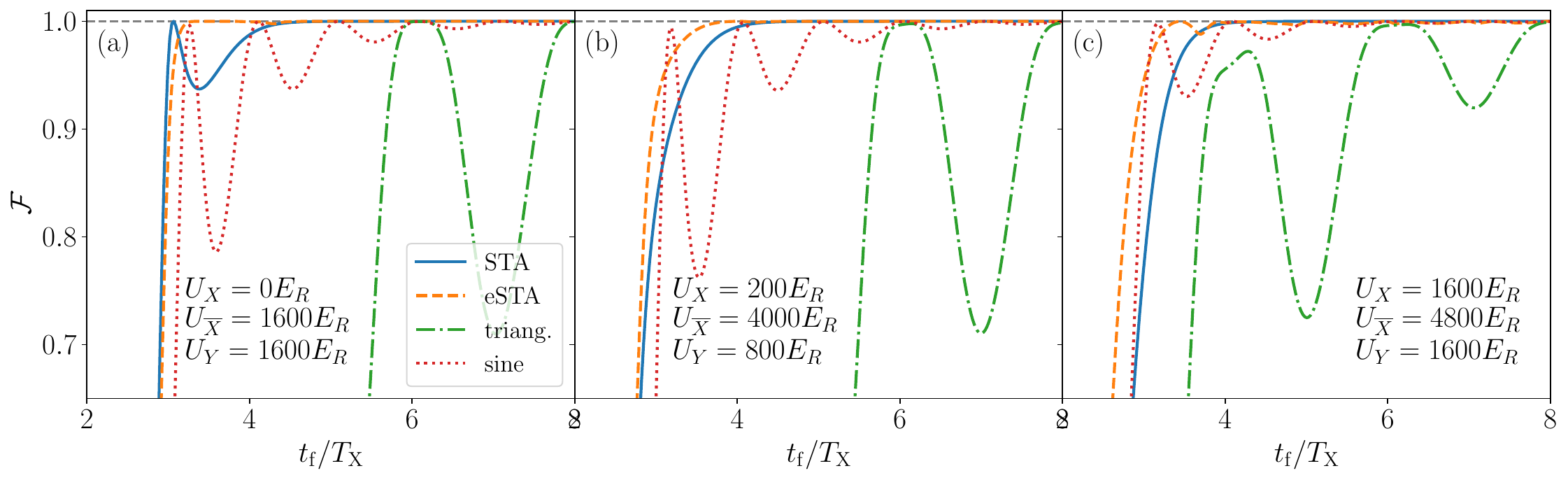}
\caption{\label{fig:FidelityResultsAll}(Color online) Dependence of the atom-transport fidelity on the transport time $t_\mathrm{f}$ for fixed potential depths using the STA- and eSTA-based transport, as well as transport based on sine- and triangular (ad-hoc) velocity profiles. The ratio of the lattice depths corresponds to that (a) the square lattice with $(U_X, U_{\overline{X}}, U_Y) = (0, 800, 800) E_R$; (b) the 1D-chains lattice with $(U_X, U_{\overline{X}}, U_Y) = (200, 4000, 800) E_R$; (c) the honeycomb lattice with $(U_X, U_{\overline{X}}, U_Y) = (1600, 4800, 1600) E_R$.
The transport distance is $d_{\mathrm{x}} = 100\,l_\mathrm{x}$ for the square and 1D-chains lattice, and  $d_{\mathrm{x}} = 50\,l_\mathrm{x}$
for the honeycomb lattice.}
\end{figure*}

In the following, we conclude our analysis of 
STA and eSTA-based atom transport by comparing the results for the atom-transport fidelity obtained using these approaches 
with those corresponding to {\em ad-hoc} (independent of the optical-lattice potential) velocity profiles. 

More specifically yet, we make comparisons 
between our STA- and eSTA results for atom transport to those based on the 
sine-shaped and triangular velocity profiles. The respective time-dependent paths $q_0^\mathrm{triang}(t)$ and
$q_0^\mathrm{sin}(t)$   
of the potential minimum corresponding to the triangular and sine-shaped velocity profiles are given by
\begin{align}
q_0^\mathrm{ triang } (t) 
&
=
\begin{cases}
v_0 \, t^2 / t_f\:, &  \; \; 0 \leq t \leq t_f/2\\
v_0  \left(2 \, t -  t_f /2 - t^2 / t_f \right)\:, &  
\; \; t_f /2 < t \leq t_f  
\end{cases}
\label{TriangProfile}\:, \\
q_0^\mathrm{ sine } (t) 
&
= \frac{ v_0 }{ 2 }\left[t - \frac{\sin\left( 2 \pi \frac{t}
{ t_f}\right)}{2 \pi}\:t_f \right] \:,
\label{SineProfile}
\end{align}
where $v_0 = 2d/t_f$ is the maximal velocity during transport over the distance $d$.

The atom-transport approach based 
on the triangular velocity profile [cf. Eq.~\eqref{TriangProfile}] is often referred to as the {\em bang-bang} approach~\cite{Torrontegui+:11,Chen+:11,Ding+:20}. This approach is accompanied by significant motional-heating effects in the fast-transport regime, a phenomenon inextricably linked with the presence of discontinuities in its corresponding acceleration profile. Consequently, such an approach leads to relatively inefficient atom transport in most practically relevant cases~\cite{Hickman+Saffman:20}. 
Somewhat better results than the ones originating from the bang-bang approach are obtained for atom transport based on the sine-shaped velocity profile [cf. Eq.~\eqref{SineProfile}]; this is intimately related to the fact that its attendant acceleration of an atom remains continuous throughout the transport process. 

By contrast to atom-transport approaches based on ad-hoc velocity profiles, the inverse-engineering 
based STA approach, and its eSTA counterpart to an even larger extent, make use of the specific form of the optical-lattice potential in question to obtain a tailored dynamical-lattice trajectory. For this reason, it is plausible to expect that these last two approaches should lead to much more 
time-efficient and robust atom transport than those based on 
pre-selected velocity profiles. 

For the optical-lattice family under consideration, a comparison between transport based on ad-hoc velocity profiles and transport based on STA- and eSTA approaches is illustrated in Fig.~\ref{fig:FidelityResultsAll} for the cases of the square-, 
1D-chains-, and honeycomb lattice.
As is evident from Fig.~\ref{fig:FidelityResultsAll},
the obtained results for the transport fidelity in the case of the triangular velocity show strong oscillations. The results obtained 
in the case of the sine-shaped velocity profile are closer -- but still inferior -- to those
originating from the STA- and 
eSTA approaches. From the results obtained using the latter approaches
it can be inferred that only one major drop in fidelity takes 
place in the case of STA-based transport in the square lattice
[cf. Fig.~\ref{fig:FidelityResultsAll}(a)], while for the remaining two lattice configurations [cf. Fig.~\ref{fig:FidelityResultsAll}(b) and (c)] such drops are completely absent for both STA 
and eSTA transport fidelities.

\section{Summary and Conclusions} \label{SummaryConclusions}
To conclude, in this paper we investigated coherent atom transport in an adjustable family of two-dimensional optical lattices which includes square, honeycomb, dimerized, and 1D-chains lattices as its special cases. We first specified concrete arrangements of acousto-optic modulators that give rise to the dynamical(moving)-lattice
effect in these systems in an arbitrary direction. We subsequently designed the corresponding dynamical-lattice trajectories using both shortcuts to adiabaticity (STA)
in the form of inverse engineering based on a 
Lewis-Riesenfeld type dynamical invariant, as well as 
their enhanced version (eSTA). Through the numerical propagation of the relevant time-dependent Schr\"{o}dinger 
equation -- using the previously obtained STA and eSTA dynamical-lattice trajectories -- we quantified the efficiency of the resulting coherent atom transport as described by the atom-transport fidelity. 

By computing transport fidelities for wide ranges of values of relevant parameters (lattice depths, transport distances, etc.) and different transport directions for four different lattice configurations, we showed 
that -- with the notable exception of the dimerized 
lattice -- the eSTA method consistently outperforms its STA counterpart; this superiority of the eSTA approach is manifested both in terms of the achievable transport times and the robustness of the resulting transport against small variations of optical-lattice depths. For the sake of completeness, we also demonstrated that our adopted eSTA approach to atom transport is far superior to alternative approaches to modeling atom transport in optical lattices, more precisely the approaches based on ad-hoc velocity profiles.

By contrast to previous theoretical investigations of coherent atom transport, based mostly on simplified scenarios, in the present work we addressed this phenomenon based on the full, anharmonic optical-lattice potential. Thus, the obtained results can be corroborated in future transport experiments in such lattices. 
Furthermore, our study also complements well the recent works on quantum-state control of Bose-Einstein condensates in optical lattices~\cite{Amri+:19}. Finally, this work may provide the impetus for further experimental implementations of STA-based control schemes for coherent atom transport in optically-trapped neutral-atom systems (including those based on the eSTA approach), the first one of which has 
been undertaken quite recently~\cite{Hwang+:25}.

\begin{acknowledgments}
V.M.S. is grateful to K. Viebahn for calling his attention to the dynamical optical-lattice family of \cite{Tarruell+:12}. This research was 
supported by the Deutsche Forschungsgemeinschaft 
(DFG) -- SFB 1119 -- 236615297.
\end{acknowledgments}

\appendix

\onecolumngrid

\section{Derivation of the expression for $G_n$} \label{DeriveGn}
In the following, we provide the derivation of an expression that serves as a prerequisite for the calculation of the scalar auxiliary function $G_{\mathbf{n}}$ [cf. Eq.~\eqref{eqExpressionG}] in the problem under consideration. For brevity, the two-dimensional index $\mathbf{n}\equiv (n_\mathrm{x},n_\mathrm{y})$ will hereafter be used whenever possible.
Furthermore, the evaluated integrals can be calculated with respect to a specific coordinate, however, they can readily be transferred to their corresponding counterparts. The excitation number of the integral indicates the coordinate over which the integration was performed.

The transport modes corresponding to the approximated Hamiltonian $H_{\mathrm{L},0}(t)$ [cf. Eq.~\eqref{approxHamiltonian}], which 
describes transport within the $x-y$ plane, have the form characteristic of a 2D harmonic oscillator.
In the coordinate representation, those transport 
modes are given by a 2D generalization of the 
expression in Eq.~\eqref{1DtransportModes}, in which 
the Lewis-Riesenfeld phase is given by Eq.~\eqref{1DLHO_LRphase}.
In other words, we have
\begin{equation}\label{eqSpatialTransportMode}
\begin{split}
\braket{\boldsymbol{r}|\Psi_\mathbf{n}}
&
=\exp \left[ -\frac{i}{\hbar} \left( E_\mathbf{n} t + \frac{m}{2} \int_0^t \mathrm{d}t'
\left[\dot{q}_\mathrm{c,x}(t')^2 + \dot{q}_\mathrm{c,y}(t')^2\right]\right) \right]\exp \left(
im \left[\dot{q}_\mathrm{c,x}(t) x + \dot{q}_\mathrm{c,y}(t) y\right]/ \hbar \right)
\\
&
\hspace{0.5cm} \times
\left(  2^n n_\mathrm{x}! n_\mathrm{y}! \pi^2 \right)^{-1/2}
\mathrm{H}_{n_\mathrm{x}} \left(\frac{x-q_{\mathrm{c}}^{\mathrm{x}}(t)}{l_\mathrm{x} \sqrt{2}} \right)
\mathrm{H}_{n_\mathrm{y}} \left(\frac{y-q_{\mathrm{c}}^{\mathrm{y}}(t)}{l_\mathrm{y} \sqrt{2}} \right)
\\
&
\hspace{0.5cm} \times
\left(2 l_\mathrm{x} \, l_\mathrm{y} \right)^{-1/2} \,
\exp \left( -\frac{[x - q_{\mathrm{c}}^{\mathrm{x}}(t)]^2}{4l_\mathrm{x}^2} \right) \,
\exp \left( -\frac{[y - q_{\mathrm{c}}^{\mathrm{y}}(t)]^2}{4l_\mathrm{y}^2} \right) \:.
\end{split}
\end{equation}
Here $E_\mathbf{n}\equiv \hbar\omega_\mathrm{x}(n_\mathrm{x} + 1/2)+ \hbar\omega_{\mathrm{y}}(n_{\mathrm{y}} + 1/2) -m a_{\mathrm{x}}^2/(2\omega_{\mathrm{x}}^2)$ is the $n$-th energy eigenvalue, which contain a constant-energy offset due to the presence of a linear term in $x$
in the approximated potential of Eq.~\eqref{eqApproxPotent}; $l_{\mathrm{j}}\equiv\sqrt
{\hbar/\left(2 m \omega_j\right)}$ is the harmonic-oscillator zero-point length along the direction $j$ ($j=x,y$), while $H_n (x)$ ($n\in \mathbb{N}$) is the Hermite polynomial~\cite{ChowBOOK:00}. 

By inserting the expression from Eq.~\eqref{eqSpatialTransportMode} 
into Eq.~\eqref{eqExpressionG}, we 
obtain the following expression for the
scalar auxiliary function:
\begin{equation}\label{eqGStart}
\begin{split}
G_\mathbf{n} =
& \int_0^{t_\mathrm{f}} \mathrm{d}t \int_{-\infty}^{\infty} \mathrm{d}Y
\int_{-\infty}^{\infty} \mathrm{d}X \, \frac{\exp\left[ i
\left( \omega_\mathrm{x} n_\mathrm{x} + \omega_\mathrm{y} n_\mathrm{y}\right)
t \right]}{\sqrt{2^n n_\mathrm{x}! n_\mathrm{y}! \pi^2}} \,
\mathrm{H}_{n_\mathrm{x}} \left[ X_\mathrm{C}(t) \right] \mathrm{H}_{n_\mathrm{y}} \left[ Y_\mathrm{C}(t)
\right] \, \exp \left[ - X_\mathrm{C}^2(t) \right] \exp
\left[ - Y_\mathrm{C}^2(t) \right]
\\
&
\times \Bigg\{
U_\mathrm{L}\left[\sqrt{2} \, X_0(t) \, l_\mathrm{x}, \sqrt{2} \, Y_0(t) \, l_\mathrm{y}\right]
-\frac{\hbar}{2}\left[ \omega_\mathrm{x} X_0^2(t) + \omega_\mathrm{y} Y_0^2(t)\right]
-
\sqrt{2} \, m \, a_\mathrm{x} \, l_\mathrm{x} \, X_0(t)
-
V_\mathrm{d,0}
\Bigg\}  \:.
\end{split}
\end{equation}
In the last equation, we introduced 
the new, dimensionless variables
$X = x / ( l_\mathrm{x} \sqrt{2} ) $ and
$Y = y / ( l_\mathrm{y} \sqrt{2} ) $,
as well as the new functions
$ X_0(t) := X - q_0^{\mathrm{x}}(t) / ( l_\mathrm{x}\sqrt{2}) $,
$ Y_0(t) := Y - q_0^{\mathrm{y}}(t) / ( l_\mathrm{y}\sqrt{2}) $,
$ X_\mathrm{C}(t) := X -  q_{\mathrm{c}}^{\mathrm{x}}(t) / (l_\mathrm{x}\sqrt{2})$ and
$ Y_\mathrm{C}(t) := Y -  q_{\mathrm{c}}^{\mathrm{y}}(t) / (l_\mathrm{y}\sqrt{2})$.

The evaluation of the integral in Eq.~\eqref{eqGStart} reduces to the calculation of several auxiliary intermediate integrals.
The first class of integrals that have to be considered are those with integrands that do not have additional dependence on $X_0$; they are given by
\begin{equation}\label{eqGInetgrationNox}
\begin{split}
W_{n_\mathrm{x}}(t)
&
= \int_{-\infty}^{\infty} \mathrm{d}X \, \exp \left[ - X_\mathrm{C}^2(t) \right]
\, \mathrm{H}_{n_\mathrm{x}}
\left[ X_\mathrm{C}(t) \right]
\\
&
= \int_{-\infty}^\infty\mathrm{d}X \sum_{m=0}^\infty\frac{1}{m!}\,\left[\frac{q_{\mathrm{c}}^{\mathrm{x}}(t)}{\sqrt{2}l_\mathrm{x} }
\right]^m
\sum_{l=0}^{n_\mathrm{x}}
\begin{pmatrix}
n_\mathrm{x}\\
l
\end{pmatrix}
\mathrm{H}_{l} \left( X \right) \left[- \sqrt{2}\:\frac{q_{\mathrm{c}}^{\mathrm{x}}(t) }{l_\mathrm{x} } \right]^{n_\mathrm{x}-l}
\mathrm{H}_{m} \left( X \right)
\exp \left( -X^2 \right) \,
\\
&
= \sqrt{\pi} \, \delta_{n_\mathrm{x},0} \:.
\end{split}
\end{equation}
This integral was evaluated by making use of the Fa\'{a} di Bruno representation of Hermite polynomials~\cite{Weisstein:21}
\begin{equation}\label{eqHermitFaadibruno}
\mathrm{H}_m(x)=(-1)^m\sum_{k_1+2k_2=m} \frac{m!}{k_1! k_2!}
(-1)^{k_1 + k_2} \left( 2 \mathnormal{x} \right)^{k_1} \:.
\end{equation}
Integrals of the form of Eq. \eqref{eqGInetgrationNox} vanish based on the well-known orthogonality condition for Hermite polynomials for all but the zero-excitation mode.
This results in constant offsets only contributing in the trivial case of $\mathbf{n}=(0,0)$,
however, for the calculation of $G_\mathbf{n}$ this case does not occur [cf. Eq.~\eqref{eqExpressionG}].

In a similar way, we can obtain the final result for an integrand that has a term linear in $X_0$.
This quantity can be straightforwardly derived using the following expression:
\begin{equation}\label{eqGInetgrationx}
\begin{split}
L_{n_\mathrm{x}}(t)
&
= \int_{-\infty}^{\infty} \mathrm{d}X \, \exp \left[ - X_\mathrm{C}^2(t) \right] \, X_0(t)
\, \mathrm{H}_{n_\mathrm{x}}
\left[ X_\mathrm{C}(t) \right]
\\
&
= \int_{-\infty}^\infty\mathrm{d}X \sum_{m=0}^\infty\frac{1}{m!}\,\left[\frac{q_{\mathrm{c}}^{\mathrm{x}}(t)}{\sqrt{2}l_\mathrm{x} }
\right]^m
\sum_{l=0}^{n_\mathrm{x}}
\begin{pmatrix}
n_\mathrm{x}\\
l
\end{pmatrix}
\mathrm{H}_{l} \left( X \right) \left[- \sqrt{2}\:\frac{q_{\mathrm{c}}^{\mathrm{x}}(t) }{l_\mathrm{x} } \right]^{n_\mathrm{x}-l}
\\
&
\hspace{0.5cm} \times \Bigg[
\frac{1}{2} \, \mathrm{H}_{m + 1} \left( X \right) - \frac{q_0^{\mathrm{x}}(t) }{\sqrt{2}\:l_\mathrm{x} }\, \mathrm{H}_{m}
\left( X \right) + m \mathrm{H}_{m - 1} \left( X \right)
\Bigg]
\exp \left( -X^2 \right) \,
\\
&
=  \sqrt{\pi} \,  \left[ \sqrt{2} \,\frac{q_{\mathrm{c}}^{\mathrm{x}}(t) }{l_\mathrm{x}}
\right]^{n_\mathrm{x}-1} 
\sum_{l=0}^{n_\mathrm{x}}
\begin{pmatrix}
n_\mathrm{x}\\
l
\end{pmatrix}
(-1)^{n_\mathrm{x} - l}
\left\{
l +
\frac{q_{\mathrm{c}}^{\mathrm{x}}(t) \left[q_{\mathrm{c}}^{\mathrm{x}}(t) - q_0^{\mathrm{x}}(t) \right]}{2 \, l_\mathrm{x}^2}
\right\}
\:.
\end{split}
\end{equation}
This result is obtained using the same steps as in the evaluation of the integral in Eq.~\eqref{eqGInetgrationNox} above, with an additional usage on the identity~\cite{ChowBOOK:00}
\begin{equation}
\label{eqRelationHermit1}
x \, \mathrm{H}_m(x)=\frac{1}{2}\:\mathrm{H}_{m+1}(x) + m \, \mathrm{H}_{m-1}(x) \:.
\end{equation}
It is straightforward to prove inductively that the only non-vanishing contributions are those with $n_\mathrm{x}\in\{0,1\}$. Therefore, the final result is given by
\begin{equation}
L_{n_\mathrm{x}}(t) = \sqrt{\pi} \left[\delta_{n_\mathrm{x},1} + \delta_{n_\mathrm{x},0} \,
\frac{q_{\mathrm{c}}^{\mathrm{x}}(t) - q_0^{\mathrm{x}}(t) }{\sqrt{2} \, l_\mathrm{x}}\right]\:.
\end{equation}

The integration involving the $X_0^2(t)$-dependent terms of Eq.~\eqref{eqGStart} entails no new methods of integration and can readily be obtained through the following
\begin{equation}\label{eqGInetgrationxSq}
\begin{split}
Q_{n_\mathrm{x}}(t)
&
= \int_{-\infty}^{\infty} \mathrm{d}X \, \exp\left[ - X_\mathrm{C}^2(t) \right]
\, X_0^2(t) \, \mathrm{H}_{n_\mathrm{x}} \left[ X_\mathrm{C}(t) \right]
\\
&
=  \int_{-\infty}^\infty \mathrm{d}X \sum_{m=0}^\infty \frac{\mathrm{H}_{m} \left( X \right)}{m!}
\left[ \frac{q_{\mathrm{c}}^{\mathrm{x}}(t)}{\sqrt{2} \, l_\mathrm{x}}\right]^m \exp \left( -X^2 \right)
\, 
\\
&
\hspace{0.5cm} \times  \left[X^2 - \sqrt{2}\: X \frac{q_0^{\mathrm{x}}(t)}{l_\mathrm{x}} +  \frac{{q_0^{\mathrm{x}}}^2(t)}{2 \,
l_\mathrm{x}^2} \right] \, \sum_{l=0}^{n_\mathrm{x}}
\begin{pmatrix}
n_\mathrm{x}\\
l
\end{pmatrix}
\mathrm{H}_{l} \left( X \right) \left[- \sqrt{2} \:\frac{q_{\mathrm{c}}^{\mathrm{x}}(t)}{l_\mathrm{x}} \right]^{n_\mathrm{x}-l}
\\
&
= \sqrt{\pi} \, \left[ \sqrt{2}
\, \frac{q_{\mathrm{c}}^{\mathrm{x}}(t)}{l_\mathrm{x}} \right]^{n_\mathrm{x}} \,
\\
&
\hspace{0.5cm} \times \sum_{l=0}^{n_\mathrm{x}}
\begin{pmatrix}
n_\mathrm{x}\\
l
\end{pmatrix}
(-1)^{n_\mathrm{x} - l}\Bigg\{\frac{1}{4} \, l \, (l-1) \, \left[ \frac{q_{\mathrm{c}}^{\mathrm{x}}(t)}{\sqrt{2}\:
l_\mathrm{x}} \right]^{-2}-l  \, q_0^{\mathrm{x}}(t) \, {q_{\mathrm{c}}^{\mathrm{x}}(t)}^{-1}+ l \Bigg\} \:,
\end{split}
\end{equation}
where use has been made of the binomial theorem. The above integral evaluates to
\begin{equation}
\begin{split}\label{eqHarmoicTreatmentDWOL1}
Q_{n_\mathbf{x}}(t)
&
= \int_{-\infty}^{\infty} \mathrm{d}X \, \exp \left[ - X_\mathrm{C}^2(t) \right]
\, X_0^2(t) \, \mathrm{H}_{n_\mathrm{x}} \left[ X_\mathrm{C}(t) \right]
\\
&
= \sqrt{\pi}  \, \left[ \delta_{n_\mathrm{x},0} \, \frac{{q_0^{\mathrm{x}}}^2(t)}{2 \: l_\mathrm{x}^2} + \delta_{n_\mathrm{x},1}
\, \frac{\sqrt{2} \, q_{\mathrm{c}}^{\mathrm{x}}(t)-q_0^{\mathrm{x}}(t)}{l_\mathrm{x}} + 2 \: \delta_{n_\mathrm{x},2}\right] \:.
\end{split}
\end{equation}

The second complex class of integrals that we need to evaluate is formed by dependents through a trigonometric relation on $X_0(t)$.
The first integral of this class that we consider is the following
\begin{equation}\label{eqIntegrationX1pre}
C_{n_\mathrm{x}}^a\left(t \right) = \int_{-\infty}^{\infty} \mathrm{d}X  \, \mathrm{H}_{n_\mathrm{x}}
\left[ X_\mathrm{C}(t) \right]  \, \exp\left[ - X_\mathrm{C}^2(t) \right]\cos \left[\sqrt{2} \, a\, k_\mathrm{L}
\, l_\mathrm{x} \, X_0(t) \right]\:,
\end{equation}
with $a\in\mathbb{R}$. This integral can be evaluated in the same manner as the previously considered ones, beginning by rewriting the cosine term in terms of exponential functions.
In this way, one obtains the expression 
\begin{eqnarray}\label{eqIntegrationX1_pre}
C_{n_\mathrm{x}}^a\left(t \right)
&=&
\frac{(-1)^{n_{\mathrm{x}}}}{2} \, \sum_{k_1+2k_2=n_\mathrm{x}} \frac{n_\mathrm{x}!}{k_1! k_2!} (-1)^{k_1 + k_2} 2^{k_1}
\sum_{l=0}^{k_1} 
\begin{pmatrix}
    k_1 \\
    l  
\end{pmatrix}
\left[- \frac{q_{\mathrm{c}}^{\mathrm{x}}(t)}{\sqrt{2} l_\mathrm{x}}\right]^{k_1 - l}
\int_{-\infty}^{\infty} \mathrm{d}X \, X^l \, \exp \left[ - \left(X - \frac{q_{\mathrm{c}}^{\mathrm{x}}(t)}{\sqrt{2} l_\mathrm{x}}\right)^2 \right] \nonumber
\\
&\times&
\Bigg\{\exp
\left[i \, \sqrt{2} \, k_\mathrm{L} \, l_\mathrm{x} \, \left(X - \frac{q_0^{\mathrm{x}}(t)}{\sqrt{2} l_\mathrm{x}}\right)\right] + \exp \left[-i\, \sqrt{2} \, k_\mathrm{L}
\, l_\mathrm{x} \, \left(X - \frac{q_0^{\mathrm{x}}(t)}{\sqrt{2} l_\mathrm{x}}\right)\right]\Bigg\} \:.
\end{eqnarray}
The integral in Eq.~\eqref{eqIntegrationX1_pre} can be computed by ordering the different terms
and employing the following general result:
\begin{equation}
\label{eqIntegralRelation}
\int_{-\infty}^\infty \mathrm{d}x \, x^n \, \exp \left( - \mathrm{a} x^2 + \mathrm{b} x +\mathrm{c} \right)
= \exp \left( \frac{b^2}{4a} + c \right) \sum_{k=0}^{\lfloor n/2 \rfloor}
\begin{pmatrix}
n \\
2k
\end{pmatrix}
\left(\mathrm{\frac{b}{2a}}\right)^{n-2k}
\frac{ \Gamma \left( k + 1/2 \right) }{\mathrm{a}^{k + 1/2}} \:,
\end{equation}
where $\Gamma(x)=\int_{0}^{\infty}t^{x-1}
\:e^{-t}\mathrm{d}t$ is the gamma function (hereafter used only
for a real-valued argument $x$)~\cite{ChowBOOK:00}.
This leads to the final expression for the cosine-dependent term 
\begin{equation}\label{eqIntegrationX1}
\begin{split}
C_{n_\mathrm{x}}^a\left(t \right)
&
= \frac{(-1)^{n_\mathrm{x}}}{2}  \sum_{k_1+2k_2=n_\mathrm{x}} \frac{n_\mathrm{x}!}{k_1! k_2!} (-1)^{k_1 + k_2} \, 2^{k_1}
\,  \exp\left(-\frac{a^2 k_\mathrm{L}^2 l_\mathrm{x}^2}{2}\right)
\,
\sum_{l=0}^{k_1}
\begin{pmatrix}
k_1\\
l
\end{pmatrix}
\left[-\frac{q_{\mathrm{c}}^{\mathrm{x}}(t)}{\sqrt{2}l_\mathrm{x}}\right]^{k_1-l}
\\
&
\hspace{0.5cm} \times
\sum_{\lambda=0}^{\lfloor l/2 \rfloor}
\begin{pmatrix}
l \\
2 \lambda
\end{pmatrix}
\Gamma \left( \lambda + 1/2 \right)
\,
\Bigg\{ \exp\left( i a \,  k_\mathrm{L} \left[ q_{\mathrm{c}}^{\mathrm{x}}(t)-q_0^{\mathrm{x}}(t) \right]
\right) \left[ \frac{q_{\mathrm{c}}^{\mathrm{x}}(t)}{\sqrt{2}\:l_\mathrm{x}} + i a \, \frac{k_\mathrm{L}
l_\mathrm{x}}{\sqrt{2}} \right]^{l-2 \lambda}
\\
&
\hspace{0.5cm}+ \exp \left( - i a \,  k_\mathrm{L} \left[ q_{\mathrm{c}}^{\mathrm{x}}(t)-q_0^{\mathrm{x}}(t) \right]
\right) \left[\frac{q_{\mathrm{c}}^{\mathrm{x}}(t)}{\sqrt{2}\:l_\mathrm{x}} - i a \, \frac{k_\mathrm{L}
l_\mathrm{x}}{ \sqrt{2} }\right]^{l-2 \lambda}\Bigg\}  \:.
\end{split}
\end{equation}

Similarly, to evaluate the sine-function counterpart of the integral in Eq.~\eqref{eqIntegrationX1_pre}, one follows analogous steps. In this manner, we arrive at the following final result:
\begin{equation}\label{eqIntegrationX115}
\begin{split}
S_{n_\mathrm{x}}^a\left(t \right)
&
=  \int_{-\infty}^{\infty} \mathrm{d}X  \, \mathrm{H}_{n_\mathrm{x}}
\left[ X_\mathrm{C}(t) \right]  \, \exp\left[ - X_\mathrm{C}^2(t) \right]\sin \left[ \sqrt{2} \, a\, k_\mathrm{L}
\, l_\mathrm{x} \, X_0(t) \right] \\
&
= (-1)^{n_\mathrm{x}-1} \frac{i}{2}  \sum_{k_1+2k_2=n_\mathrm{x}} \frac{n_\mathrm{x}!}{k_1! k_2!} (-1)^{k_1 + k_2} \, 2^{k_1}
\,  \exp\left(-\frac{a^2 k_\mathrm{L}^2 l_\mathrm{x}^2}{2}\right)
\,
\sum_{l=0}^{k_1}
\begin{pmatrix}
k_1\\
l
\end{pmatrix}
\left[-\frac{q_{\mathrm{c}}^{\mathrm{x}}(t)}{\sqrt{2}l_\mathrm{x}}\right]^{k_1-l}
\\
&
\hspace{0.5cm} \times
\sum_{\lambda=0}^{\lfloor l/2 \rfloor}
\begin{pmatrix}
l \\
2 \lambda
\end{pmatrix}
\Gamma \left( \lambda + 1/2 \right)
\,
\Bigg\{ \exp\left( i a \,  k_\mathrm{L} \left[ q_{\mathrm{c}}^{\mathrm{x}}(t)-q_0^{\mathrm{x}}(t) \right]
\right) \left[ \frac{q_{\mathrm{c}}^{\mathrm{x}}(t)}{\sqrt{2}\:l_\mathrm{x}} + i a \, \frac{k_\mathrm{L}
l_\mathrm{x}}{\sqrt{2}} \right]^{l-2 \lambda}
\\
&
\hspace{0.5cm} - \exp \left( - i a \,  k_\mathrm{L} \left[ q_{\mathrm{c}}^{\mathrm{x}}(t)-q_0^{\mathrm{x}}(t) \right]
\right) \left[\frac{q_{\mathrm{c}}^{\mathrm{x}}(t)}{\sqrt{2}\:l_\mathrm{x}} - i a \, \frac{k_\mathrm{L}
l_\mathrm{x}}{ \sqrt{2} }\right]^{l-2 \lambda}\Bigg\}  \:.
\end{split}
\end{equation}
It should be noticed that the only differences of this last result compared to Eq.~\eqref{eqIntegrationX1} are 
the change of sign in the curly brackets and the additional prefactor of $(-i)$.

Another term that has to be evaluated is similar 
to Eq.~\eqref{eqIntegrationX1pre}, but instead of 
a simple cosine term, it contains a cosine-squared,
term. This integral is given by
\begin{eqnarray}\label{eqIntegrationC2Approx}
\widetilde{C}_{n_\mathrm{x}}\left(t \right)
&=&
\int_{-\infty}^{\infty} \mathrm{d}X \, \mathrm{H}_{n_\mathrm{x}}\left[ X_\mathrm{C}(t) \right] \, \exp
\left[ - X^2_\mathrm{C}(t) \right]  \cos^2\left[ \sqrt{2} \,  k_\mathrm{L}  \, l_\mathrm{x} \, X_0(t) \right] \nonumber
\\
&=&
\frac{1}{4} \, \sum_{k_1+2k_2=n_\mathrm{x}} \frac{n_\mathrm{x}!}{k_1! k_2!} (-1)^{k_1 + k_2} 2^{k_1}
\sum_{l=0}^{k_1} 
\begin{pmatrix}
    k_1 \\
    l  
\end{pmatrix}
\left[- \frac{q_{\mathrm{c}}^{\mathrm{x}}(t)}{\sqrt{2} l_\mathrm{x}}\right]^{k_1 - l}
\int_{-\infty}^{\infty} \mathrm{d}X \, X^l \, \exp \left[ - \left(X - \frac{q_{\mathrm{c}}^{\mathrm{x}}(t)}{\sqrt{2} l_\mathrm{x}}\right)^2 \right] \nonumber
\\
&\times&
\Bigg\{\exp
\left[i \, 2^{3/2} \, k_\mathrm{L} \, l_\mathrm{x} \, \left(X - \frac{q_0^{\mathrm{x}}(t)}{\sqrt{2} l_\mathrm{x}}\right)\right] + \exp \left[-i\, 2^{3/2} \, k_\mathrm{L}
\, l_\mathrm{x} \, \left(X - \frac{q_0^{\mathrm{x}}(t)}{\sqrt{2} l_\mathrm{x}}\right)\right]+ 2 \Bigg\} \:.
\end{eqnarray}

Since no new techniques were used to evaluate this integral,
we just state the final result,
which is given by
\begin{equation}\label{eqIntegrationC2ApproxFinal}
\begin{split}
\widetilde{C}_{n_\mathrm{x}}\left(t \right)
&
= \frac{(-1)^{n_\mathrm{x}}}{4} \, \sum_{k_1 + 2 k_2 = n_\mathrm{x}} \frac{n_\mathrm{x}!}
{k_1! k_2!} (-1)^{k_1 + k_2}2^{k_1}\,
\sum_{l=0}^{k_1}
\begin{pmatrix}
    k_1\\
    l
\end{pmatrix}
\left[- \frac{q_{\mathrm{c}}^{\mathrm{x}}(t)}{\sqrt{2} l_\mathrm{x}}\right]^{k_1 - l} \,
\sum_{\lambda=0}^{\lfloor l/2 \rfloor}
\begin{pmatrix}
l \\
2 \lambda
\end{pmatrix}
\Gamma \left( \lambda + 1/2 \right) \, \exp\left(- 2\, k_\mathrm{L}^2 \, l_\mathrm{x}^2 \right)
\\
&
\times \Bigg\{
\left[\frac{q_{\mathrm{c}}^{\mathrm{x}}(t)}{\sqrt{2} l_\mathrm{x}} + i \, \sqrt{2} \, k_\mathrm{L} \, l_\mathrm{x} \right]^{l - 2 \lambda}
\exp\Big\{
+i \, 2 \, k_\mathrm{L} \left[q_{\mathrm{c}}^{\mathrm{x}}(t) -
q_0^{\mathrm{x}}(t)\right]
\Big\}
\\
&
+ \hspace{0.4cm}
\left[\frac{q_{\mathrm{c}}^{\mathrm{x}}(t)}{\sqrt{2} l_\mathrm{x}} - i \, \sqrt{2} \, k_\mathrm{L} \, l_\mathrm{x}\right]^{l - 2 \lambda}
\exp\Big\{
-i \, 2 \, k_\mathrm{L} \left[q_{\mathrm{c}}^{\mathrm{x}}(t) -
q_0^{\mathrm{x}}(t)\right]
\Big\}
\Bigg\} + \frac{\sqrt{\pi}}{2}\, \delta_{n_\mathrm{x},0}\:.
\end{split}
\end{equation}

The same holds true for the sine-counterpart of the above integral, 
which results in an almost identical expression:
\begin{equation}\label{eqIntegrationS2Approx}
\begin{split}
\widetilde{S}_{n_\mathrm{x}}\left( t \right)
&=
\int_{-\infty}^{\infty} \mathrm{d}X \, \mathrm{H}_{n_\mathrm{x}}\left[ X_\mathrm{C}(t) \right] \, \exp
\left[ - X_\mathrm{C}(t)^2 \right]  \sin^2\left[ \sqrt{2} \,  k_\mathrm{L}  \, l_\mathrm{x} \, X_0(t) \right]
\\
&
= \frac{(-1)^{n_\mathrm{x}+1}}{4} \, \sum_{k_1 + 2 k_2 = n_\mathrm{x}} \frac{n_\mathrm{x}!}
{k_1! k_2!} (-1)^{k_1 + k_2}2^{k_1}\,
\sum_{l=0}^{k_1}
\begin{pmatrix}
    k_1\\
    l
\end{pmatrix}
\left[- \frac{q_{\mathrm{c}}^{\mathrm{x}}(t)}{\sqrt{2} l_\mathrm{x}}\right]^{k_1 - l} \,
\sum_{\lambda=0}^{\lfloor l/2 \rfloor}
\begin{pmatrix}
l \\
2 \lambda
\end{pmatrix}
\Gamma \left( \lambda + 1/2 \right) \, \exp\left(- 2\, k_\mathrm{L}^2 \, l_\mathrm{x}^2 \right)
\\
&
\times \Bigg\{
\left[\frac{q_{\mathrm{c}}^{\mathrm{x}}(t)}{\sqrt{2} l_\mathrm{x}} + i \, \sqrt{2} \, k_\mathrm{L} \, l_\mathrm{x} \right]^{l - 2 \lambda}
\exp\Big\{
+i \, 2 \, k_\mathrm{L} \left[q_{\mathrm{c}}^{\mathrm{x}}(t) -
q_0^{\mathrm{x}}(t)\right]
\Big\}
\\
&
+ \hspace{0.4cm}
\left[\frac{q_{\mathrm{c}}^{\mathrm{x}}(t)}{\sqrt{2} l_\mathrm{x}} - i \, \sqrt{2} \, k_\mathrm{L} \, l_\mathrm{x}\right]^{l - 2 \lambda}
\exp\Big\{
-i \, 2 \, k_\mathrm{L} \left[q_{\mathrm{c}}^{\mathrm{x}}(t) -
q_0^{\mathrm{x}}(t)\right]
\Big\}
\Bigg\} + \frac{\sqrt{\pi}}{2}\, \delta_{n_\mathrm{x},0}\:.
\end{split}
\end{equation}

By combining the obtained results for the integrals discussed so far and making use of the orthogonality of Hermite polynomials, the scalar auxiliary function $G_\mathbf{n}$ can be reduced to the time integral
\begin{equation}\label{eqGEnd1}
\begin{split}
G_\mathbf{n}
=
&
-\int_0^{t_\mathrm{f}} \mathrm{d}t \, \frac{\exp \left[ i \left( \omega_\mathrm{x}
n_\mathrm{x} + \omega_\mathrm{y} n_\mathrm{y}\right) t \right]}{\sqrt{2^n n_\mathrm{x}!
n_\mathrm{y}! \pi}}
\Bigg\{
\eta_{0,\boldsymbol{n}}(t)
+
\eta_{1,\boldsymbol{n}}(t)
+
\eta_{2,\boldsymbol{n}}(t)
\\
&
+\hbar \sqrt{\pi} \Bigg[ \omega_\mathrm{x} \delta_{n_\mathrm{y},0} \,
\left(\delta_{n_\mathrm{x},1}  \, \frac{q_{\mathrm{c}}^{\mathrm{x}}(t)-q_0^{\mathrm{x}}(t)}{\sqrt{2} l_\mathrm{x}} + \delta_{n_\mathrm{x},2}
\right)+ \omega_\mathrm{y} \delta_{n_\mathrm{x},0} \, \left(\delta_{n_\mathrm{y},1} \,
\frac{q_{\mathrm{c}}^{\mathrm{y}}(t)-q_0^{\mathrm{y}}(t)}{\sqrt{2} l_\mathrm{y}} + \delta_{n_\mathrm{y},2}\right)\Bigg]
\\
&
+
\sqrt{2 \, \pi} \, m \, a_\mathrm{x} \, l_\mathrm{x} \, \delta_{n_\mathrm{x},1}
\, \delta_{n_\mathrm{y},0} 
\Bigg\} \:,
\end{split}
\end{equation}
where $\eta_{i,\boldsymbol{n}}(t)$, $i\in\left\{0, 1, 2 \right\}$ stand for the following functions of time:

\begin{equation}\label{eqGEndfunc1}
\begin{split}
\eta_{0,\boldsymbol{n}}(t)
=
&
\:U_X \, \widetilde{S}_{n_\mathrm{x}}(t) \, \delta_{n_\mathrm{y},0} 
+ U_Y \, \widetilde{C}_{n_\mathrm{y}}(t) \, \delta_{n_\mathrm{x},0} \:,
\\
\eta_{1,\boldsymbol{n}}(t)
=
&  \:2\gamma\:\sqrt{\frac{U_X \, U_Y}{\pi}}\:
S_{n_\mathrm{x}}^{a=1}(t) \, C_{n_\mathrm{y}}^{a=1}(t)\:\cos\phi  \:,
\\
\eta_{2,\boldsymbol{n}}(t) 
=
&
\:\frac{U_{\widebar{X}}}{2}
\left[
C_{n_\mathrm{x}}^{a=2}(t)\:\cos\theta
- S_{n_\mathrm{x}}^{a=2}(t)\:\sin\theta
\right]  \, \delta_{n_\mathrm{y},0}  \: .
\end{split}
\end{equation}
The integration over time in Eq.~\eqref{eqGEnd1}, the final step in the evaluation of $G_\mathbf{n}$, has to be carried out numerically.

\section{Derivation of the expression for $\mathbf{K}_\mathbf{n}$}\label{DeriveKn}
In the following, we derive an expression that can be used as the basis for a numerical evaluation of the second (vector) auxiliary
function $\mathbf{K}_{\mathbf{n}}$ [cf. Eq.~\eqref{eqExpressionK}] in the problem at hand.

To derive the desired expression for $\mathbf{K}_{\mathbf{n}}$, the gradient with respect to $\boldsymbol{\zeta}$,
and subsequently to $\boldsymbol{\alpha}$, of the Hamiltonian $H_\mathrm{DW}$ is needed. Since the
differentiations are straightforward, we just state the resulting equation for $\mathbf{K}_{\mathbf{n}}$:
\begin{equation}\label{eqKDWOLafterZ}
\begin{split}
\mathbf{K}_\mathbf{n}
&
=  \, \int_0^{t_\mathrm{f}} \, \mathrm{d}t \, \, \int_{-\infty}^\infty \, \mathrm{d}Y \,
\int_{-\infty}^\infty \, \mathrm{d}X
\frac{ 1  }{\sqrt{2^n n_\mathrm{x}! n_\mathrm{y}! \pi^2}} \exp \left[ i
\left( n_\mathrm{x} \omega_\mathrm{x} + n_\mathrm{y} \omega_\mathrm{y} \right) t \right]
\\
&
\hspace{0.5cm} \times  \mathrm{H}_{n_\mathrm{x}} \left[ X_\mathrm{C}(t) \right] \mathrm{H}_{n_\mathrm{y}} \left[
Y_\mathrm{C}(t) \right] \, \exp \left[ - X_\mathrm{C}^2(t) \right]
\exp \left[ - Y_\mathrm{C}^2(t) \right] 
\\
&
\hspace{0.5cm} \times
\boldsymbol{\nabla}_{\boldsymbol{\alpha}}
\underbrace{
\left\{
U_X [\sqrt{2}\:X_0(t)l_{\mathrm{x}}] +
U_Y [\sqrt{2}\:Y_0(t)l_{\mathrm{y}}] +
U_{XY}[\sqrt{2}\:X_0(t)l_{\mathrm{x}}, \sqrt{2}\:Y_0(t)l_{\mathrm{y}}]
\right\}}_{U_\mathrm{D}}
\Big|_{\boldsymbol{\alpha}=\boldsymbol{0}}
\:.
\end{split}
\end{equation}
The derivations of the potential contributions are given by
\begin{align}
    \boldsymbol{\nabla}_{\boldsymbol{\alpha}} U_X 
    [x-q_0^{\mathrm{x}}(\alpha)]
    &=
    \:\:\: U_X k_\mathrm{L} \sin\left[2 k_\mathrm{L} (x-q_0^{\mathrm{x}})\right] \boldsymbol{\nabla}_{\boldsymbol{\alpha}} f_{\mathrm{x}}
    - U_{\widebar{X}} k_\mathrm{L} \sin\left[2 k_\mathrm{L} (x-q_0^{\mathrm{x}}) + \theta\right] \boldsymbol{\nabla}_{\boldsymbol{\alpha}} f_{\mathrm{x}} \:, \nonumber
    \\
    \boldsymbol{\nabla}_{\boldsymbol{\alpha}} U_Y [y-q_0^{\mathrm{y}}(\alpha)]
    &=
    - U_Y k_\mathrm{L} \sin\left[2 k_\mathrm{L} (y-q_0^{\mathrm{y}})\right] \boldsymbol{\nabla}_{\boldsymbol{\alpha}} f_{\mathrm{y}} 
    \:, \nonumber
    \\
    \boldsymbol{\nabla}_{\boldsymbol{\alpha}} U_{XY}[x-q_0^{\mathrm{x}}(\alpha), y-q_0^{\mathrm{y}}(\alpha)]
    &=
    - 2 \, k_\mathrm{L} \, \gamma \, \cos \phi \, \sqrt{U_X U_Y}
    \Big\{
    \sin\left[k_\mathrm{L}(x-q_0^{\mathrm{x}})\right] \, \sin\left[k_\mathrm{L}(y-q_0^{\mathrm{y}})\right]
    \boldsymbol{\nabla}_{\boldsymbol{\alpha}} f_{\mathrm{y}} \nonumber\\
    & \hspace{3.5cm}-
    \cos\left[k_\mathrm{L}(x-q_0^{\mathrm{x}})\right] \, \cos\left[k_\mathrm{L}(y-q_0^{\mathrm{y}})\right]
    \boldsymbol{\nabla}_{\boldsymbol{\alpha}} f_{\mathrm{x}}
    \Big\} \:.
\end{align}

In fact, all of the needed integrals have already been 
carried out in the evaluation of the first auxiliary function (cf. Appendix~\ref{DeriveGn}).
Thus, here we just state the final result
\begin{equation}
\begin{split}
\mathbf{K}_\mathbf{n}
&
= \int_0^{t_\mathrm{f}} \, \mathrm{d}t \,
\frac{\exp \left[ i
\left( n_\mathrm{x} \omega_\mathrm{x} + n_\mathrm{y} \omega_\mathrm{y} \right) t \right]}
{\sqrt{2^n n_\mathrm{x}! n_\mathrm{y}! \pi}}
\Big[ 
\tilde{\boldsymbol{\eta}}_{0,\boldsymbol{n}}(t) \, \delta_{n_\mathrm{y}, 0} +
\tilde{\boldsymbol{\eta}}_{1,\boldsymbol{n}}(t) \, \delta_{n_\mathrm{x}, 0}+
\tilde{\boldsymbol{\eta}}_{2,\boldsymbol{n}}(t)
\Big] \:,
\end{split}
\end{equation}
where $\tilde{\boldsymbol{\eta}}_{0,\boldsymbol{n}}(t)$, $\tilde{\boldsymbol{\eta}}_{1,\boldsymbol{n}}(t)$ and
$\tilde{\boldsymbol{\eta}}_{2,\boldsymbol{n}}(t)$
are vector functions of time given by:
\begin{equation}\label{eqGEndDWOL2}
\begin{split}
\tilde{\boldsymbol{\eta}}_{0,\boldsymbol{n}}(t)
&=
\:\:\:U_X \,  k_\mathrm{L} \, S_{n_\mathrm{x}}^{a=2}(t)
\boldsymbol{\nabla}_{\boldsymbol{\alpha}} f_{\mathrm{x}}
-U_{\widebar{X}} \,  k_\mathrm{L} \,
\left[
S_{n_\mathrm{x}}^{a=2}(t) \cos \theta +
C_{n_\mathrm{x}}^{a=2}(t) \sin \theta
\right]
\boldsymbol{\nabla}_{\boldsymbol{\alpha}} f_{\mathrm{x}} \:,
\\
\tilde{\boldsymbol{\eta}}_{1,\boldsymbol{n}}(t)
&=
-U_Y \,  k_\mathrm{L} \, S_{n_\mathrm{y}}^{a=2}(t)
\boldsymbol{\nabla}_{\boldsymbol{\alpha}} f_{\mathrm{y}} \:,
\\
\tilde{\boldsymbol{\eta}}_{2,\boldsymbol{n}}(t)
&=
-2 \, k_\mathrm{L} \, \gamma \, \cos \phi \, \sqrt{\frac{U_X U_Y}{\pi}}
\Big[
S_{n_\mathrm{x}}^{a=1}(t)
S_{n_\mathrm{y}}^{a=1}(t)
\boldsymbol{\nabla}_{\boldsymbol{\alpha}} f_{\mathrm{y}} 
-
C_{n_\mathrm{x}}^{a=1}(t)
C_{n_\mathrm{y}}^{a=1}(t)
\boldsymbol{\nabla}_{\boldsymbol{\alpha}} f_{\mathrm{x}}
\Big]
\: .
\end{split}
\end{equation}
The remaining integral over time can only be evaluated numerically, as has already been done in the evaluation of the first auxiliary function $G_\mathbf{n}$ [cf. Eq.~\eqref{eSTAbasics}] in Appendix~\ref{DeriveGn}.

\addvspace{2\baselineskip}

\twocolumngrid
\section{FSOM-based numerical propagation of TDSEs } \label{ReviewFSOM}
In the following, we briefly review the essential ingredients 
of the use of the FSOM for solving TDSEs.

Let us consider Cauchy-type initial-value problems $\partial_t f(\mathbf{r},t)=\hat{A}(\mathbf{r},t)f(\mathbf{r},t)$,
where $\hat{A}(\mathbf{r},t)$ is an operator and the initial condition is given by $f(\mathbf{r},t)=
f_0(\mathbf{r})$. The FSOM 
is the method of choice for solving such problems in cases where $\hat{A}$ can be written as a sum of 
an operator that can straightforwardly be diagonalized in real space and another one 
that can be diagonalized in Fourier space. One of the most common examples of such problems is the TDSE~\cite{Feit+:82,Bandrauk+Shen:93}. The crucial idea facilitating the use of FSOM for solving TDSEs is to approximate the time-evolution operator of the system under consideration by a product of operators that are either diagonal in real space or in Fourier space.

Consider the TDSE that describes the motion of a particle of mass $m$ in the potential $U(\vec{r})$. The underlying single-particle Hamiltonian is given by
\begin{equation}\label{SingleAtomHamiltonian}
H=-\frac{\hbar^2\nabla^2}{2m}+U(\vec{r})\:.
\end{equation}
To approximate the time-evolution operator that corresponds to this last Hamiltonian we invoke the identity (known as the symmetrized Trotter decomposition)~\cite{GilmoreBOOK:12}
\begin{equation}\label{SymmetricTrotter}
e^{i(A+B)\delta t}=e^{iA\delta t/2}e^{iB\delta t}e^{iA\delta t/2}
+\mathcal{O}(\delta t^3)\:,
\end{equation}
which holds for arbitrary Hermitian operators $A$ and $B$. In the special case $A=U(\vec{r})$, $B=-\hbar^2\nabla^2/(2m)$ of this last identity, we obtain an accurate second-order time-stepping scheme 
~\cite{Pechukas+Light:66}
\begin{eqnarray}\label{eqFSOMBCH}
\Psi(\vec{r},t&+&\delta t) = \exp\left[-\frac{i}{\hbar}\:U(\vec{r})
\frac{\delta t}{2}\right]\exp\left(i\frac{\hbar\nabla^2}{2m}\:\delta t\right) \nonumber \\
&\times& \exp\left[-\frac{i}{\hbar}\:U(\vec{r})\frac{\delta t}{2}\right]
\Psi(\vec{r},t)+ \mathcal{O}(\delta t^3) 
\end{eqnarray}
for the propagation of the wave-function 
$\Psi(\vec{r},t)$ of a system described 
by the Hamiltonian in Eq.~\eqref{SingleAtomHamiltonian}.
This approach to evaluating the time evolution -- in which the potential part of the total Hamiltonian is applied for a half time step before and after the kinetic part -- is often referred to as symmetric splitting (or, alternatively, as Strang splitting~\cite{Speth+:2013}) and its principal advantage compared to the conventional approach is that it reduces the error to $\mathcal{O}(\delta t^3)$.

The form of Eq.~\eqref{eqFSOMBCH} permits independent treatment of the different exponential terms. Therefore, it is pertinent to Fourier-transform the kinetic term, recasting at 
the same time the RHS of Eq.~\eqref{eqFSOMBCH} using the identity
\begin{eqnarray}\label{eqFourierRepresentation}
&&\exp\left(i\frac{\hbar\nabla^2}{2m}\:\delta t\right)\exp
\left[-\frac{i}{\hbar}\:U(\vec{r})\frac{\delta t}{2}\right]\Psi(\vec{r},t)  \\
&=& F^{-1}\left[e^{-i\frac{\hbar k^2}{2m}\delta t}\:F\left\{\exp\left[-\frac{i}{\hbar}
\:U(\vec{r})\frac{\delta t}{2}\right]\Psi(\vec{r},t)\right\}\right] \nonumber \:,
\end{eqnarray}
where $F[\ldots]$ stands for the Fourier transform of the argument and $F^{-1}[\ldots]$ for its inverse.

By successively applying the time-propagation scheme based on Eq.~\eqref{eqFSOMBCH} $N_t$ times on an initial wave-function $\Psi(\vec{r},t)$ one obtains
a numerical solution of a TDSE at time $t'=t+N_t\delta t$. In actual numerical implementations of the FSOM, the wave-function is represented on a discrete rectangular spatial grid with $N_s$ points and the 
exact (continuous) Fourier transform is
approximated by its discrete counterpart. 
The computational burden involved in propagating 
the wave function is dominated by the need to compute its spatial Fourier transformation and its inverse.
If carried out using the fast Fourier transform (FFT) algorithm~\cite{NRcBook}, an elementary step in these transformations entails $\mathcal{O}(N_s\log_2 N_s)$
operations.

\section{Ground-state calculation via the ITE approach} \label{FSOMgroundState}
Here we provide a short introduction into the ITE approach, which established itself as 
one of the most robust computational methods for obtaining ground 
states of quantum systems~\cite{Anderson:75,Feit+:82}.
Its relationship to the FSOMa - its counterpart utilized for computing dynamics - bears close analogy to the connection between Lanczos-type diagonalizations~\cite{Stojanovic:20} and Chebyshev-propagator methods~\cite{Stojanovic+Salom:19}.
It is prudent to start our review of the ITE approach using the abstract, representation-independent notation for quantum states.

Consider the problem of obtaining the ground
state of a Hamiltonian denoted by $H$ in the following. One starts by expanding
a quantum state $|\phi\rangle$ in terms of the eigenstates $\{|\psi_n\rangle,\:n=0,1,\ldots\}$
of $H$, i.e. $|\phi\rangle=\sum_{n}c_n|\psi_n\rangle$, assuming that the sought-after ground state $|\psi_0\rangle$ has a nonvanishing contribution to this expansion
(i.e. $c_0 \equiv \langle
\psi_0|\phi\rangle \neq 0$). Assuming that 
$|\phi\rangle$ is the initial ($t=0$)
state of the system, its states 
$|\phi(t)\rangle\equiv e^{-iHt/\hbar}|\phi\rangle$ at later 
times $t>0$ are given by the linear combinations 
\begin{equation}\label{eqRealTimeEvolution}
|\phi(t)\rangle=\sum_{n} c_n\: e^{-iE_n t/\hbar}|\psi_n\rangle
\end{equation}
of the time-evolved eigenstates $e^{-iE_n t/\hbar}|\psi_n\rangle$ of $H$, where $E_n$ is the eigenvalue of $H$ that corresponds 
to the eigenstate $|\psi_n\rangle$.

By performing a Wick rotation into the complex plane (i.e. by making the substitution $t\rightarrow \tau=it$), Eq.~\eqref{eqRealTimeEvolution} goes over into 
\begin{equation}\label{eqImagTimeEvolution}
|\phi(\tau)\rangle=\sum_{n}c_n\: e^{-E_n\tau/\hbar}|\psi_n\rangle\:.
\end{equation}
Therefore, $|\phi(\tau)\rangle$ is an exponentially-decaying superposition of
the eigenstates $|\psi_n\rangle$ with the decay rates given by the energy eigenvalues
$E_n$. Given that the decay rate of the ground state is the smallest one, in the large-$\tau$
limit one arrives at
\begin{equation}\label{LargeTauEvolution}
|\phi(\tau)\rangle 
\approx c_0\:e^{-E_0\tau/\hbar}|\psi_0\rangle\:.
\end{equation}
Therefore, given a trial state $|\phi(\tau=0)\rangle\equiv|\phi\rangle$ with a nonvanishing overlap with the sought-after ground state, the ITE approach yields a monotonously decreasing functional 
of $\tau$ that converges to the ground-state energy $E_0$ of the system as 
$\tau\rightarrow\infty$, i.e.
\begin{equation}\label{GSenergy}
E_0 = \lim_{\tau\rightarrow\infty}
\langle\phi(\tau)|\:H\:|\phi(\tau)\rangle\:.
\end{equation}

In the atom-transport problem under consideration (cf. Sec.~\ref{SingleAtomTransport}) -- and many other problems -- one makes use of the coordinate representation, in which a quantum state $|\psi\rangle$ is represented by
the wave function $\psi(\mathbf{r})\equiv\langle \mathbf{r}|\psi\rangle$ (with $|\mathbf{r}\rangle$ being the
eigenstates of the position operator). The counterpart of Eq.~\eqref{eqImagTimeEvolution} in this representation reads
\begin{equation}\label{eqImagTimeCoordinate}
\phi(\vec{r},\tau)=\sum_{n} c_n e^{-E_n\tau/\hbar}\psi_n(\vec{r})\:,
\end{equation}
where $\phi(\vec{r},\tau)\equiv \langle\mathbf{r}|\phi(\tau)\rangle$ and $\psi_n(\vec{r})
\equiv\langle\mathbf{r}|\psi_n\rangle$ are the relevant wave functions. The remaining steps in the application of the ITE approach within this representation are carried out in complete analogy with the above representation-independent expressions \eqref{LargeTauEvolution}
and \eqref{GSenergy}.

\section{Derivation of dimensionless units used in the numerical simulation} \label{DerivationSimulation}
To facilitate an accurate numerical simulation of atom transport, it is 
pertinent to introduce dimensionless units derived from the intrinsic parameters of the system. This new choice of units leads to improved numerical stability and avoids potential underflow- or overflow issues that may arise when working directly with physical units. In particular, we define characteristic time-, length-, and energy scales based on the harmonic reference frequency of the system, which are then 
used consistently throughout the simulation. More specifically
yet, we set:
\begin{align}
t_{\mathrm{sys}} \equiv \omega_{\mathrm{x}}^{-1}\:, \hspace{0.5cm}
l_{\mathrm{sys}} \equiv \sqrt{2}\:l_{\mathrm{x}}\:,  \hspace{0.5cm}
E_{\mathrm{sys}} \equiv \hbar \omega_{\mathrm{x}}\:.\nonumber
\end{align}
In particular, the advantage of including the additional 
factor of $\sqrt{2}$ in the reference length of the approximate Hamiltonian is that it implicitly rescales the kinetic term such that the effective mass is set to unity, rather than $1/2$; this choice eliminates unnecessary prefactors in the Hamiltonian and leads to a simpler and more transparent numerical implementation.

With the above definitions in place, the energy of the system can be expressed as
\begin{equation}
U_j = \tilde{U}_j E_{\mathrm{sys}} \equiv U_j^R E_{\mathrm{R}} \qquad (\:j\in\{X, Y, \widebar{X}\}\:) \:.
\end{equation}
In our simulations, we specify the energy scales in terms of the recoil energy $E_{\mathrm{R}}$ 
[cf. Sec.~\ref{SingleAtomTransport}] by setting $U_j^R$ externally, while the numerical implementation 
relies on the corresponding dimensionless harmonic reference units $\tilde{U}_X$.
However, being defined through the wavenumber $k_{\mathrm{L}}$, the recoil energy depends explicitly on the characteristic length scale of the 
system, resulting in a nontrivial connection between the energy scales.

By expressing the wavenumber $k_{\mathrm{L}}$ in system units as $k_{\mathrm{L}} = \tilde{k}/l_{\mathrm{sys}}$, the recoil energy 
becomes directly related to the dimensionless wavenumber via
\begin{equation}\label{eqRecoilEnergySU}
\frac{\hbar^2 k_{\mathrm{L}}^2}{2m} = \frac{\hbar^2 \tilde{k}^2 }{2 m l_{\mathrm{sys}}^2} 
\equiv\frac{\tilde{k}^2 }{2}\:\hbar \omega_{\mathrm{x}}\:.
\end{equation}
Further, by making use of the 
reference timescale $t_{\mathrm{sys}} = \omega_{\mathrm{x}}^{-1}$, together 
with the derived harmonic frequency of the approximated potential [cf. Eq.~\eqref{eqHarmonicFreuency}], 
we obtain
\begin{align}\label{equatomegax}
\omega_{\mathrm{x}}^2
&= \frac{2 U_X k_{\mathrm{L}}^2}{m}
\:g_{\mathrm{x}} \nonumber \\
&= 2\tilde{k}^2 \tilde{U}_X g_{\mathrm{x}}  \frac{\hbar \omega_{\mathrm{x}}}{l_{sys}^2 m} \\
&= 2\tilde{k}^2 \tilde{U}_X g_{\mathrm{x}}\: \omega_{\mathrm{x}}^2 \:, \nonumber
\end{align}
where we define the geometric factor 
in the $x$-direction as
\begin{align}
g_{\mathrm{x}} \equiv \left(1-\frac{U_{\widebar{X}}}{U_X}\right) \cos^2 (k_{\mathrm{L}}x_0) \:.
\end{align}
It is worthwhile noting that in 
Eq.~\eqref{equatomegax} above $\omega_x^2$ 
enters multiplicatively on both sides, thus 
implying that the prefactor on the RHS must 
equal unity. This is a direct consequence of defining the system variables in terms of the effective frequency corresponding to the 
harmonically-approximated optical potential.
As a result, one obtains a simple relation between the wavenumber and the potential depth in system units:
\begin{align}
\tilde{k} = \frac{1}{\sqrt{2\tilde{U}_X g_{\mathrm{x}}}} \:.
\end{align}
Along with the previously derived relation for the recoil energy in system units [cf. Eq.~\eqref{eqRecoilEnergySU}], we obtain
\begin{equation}
\tilde{U}_X = U_X^R\: \frac{E_R}{E_{\mathrm{sys}}} \equiv 
 U_X^R\:\frac{\hbar^2 \tilde{k}^2}{2 m 
 \:\hbar \omega_{\mathrm{x}} l_{\mathrm{sys}}^2}  \:,
 \end{equation}
which further leads to 
\begin{equation} 
\tilde{U}_X = U_X^R \:\frac{\tilde{k}^2}{2}
= U_X^R \:\frac{1}{4 \tilde{U}_X g_{\mathrm{x}}}\:.
\end{equation}
By solving the last equation 
for $\tilde{U}_X$, we arrive at the
sought-after relation between the potential depth in terms of system 
and recoil energies in the $x$ direction:
\begin{align}
\tilde{U}_X = \sqrt{\frac{U_X^R}{4 g_{\mathrm{x}}}}.
\end{align}
Analogous relations hold for the two remaining potential depths, i.e.
\begin{equation}\label{EqUYUXbar}
\tilde{U}_Y = \sqrt{\frac{U_Y^R \kappa_Y}{4 g_{\mathrm{x}}}}\:, \quad
\tilde{U}_{\widebar{X}} = \sqrt{\frac{U_{\widebar{X}}^R \kappa_{\widebar{X}}}{4 g_{\mathrm{x}}}} \:,
\end{equation}
where, in order to further simplify these last equations, we introduced the relative potential depths:
\begin{align}
\kappa_Y = \frac{U_Y}{U_X}\:, \hspace{0.5cm}
\kappa_{\widebar{X}} = \frac{U_{\widebar{X}}}{U_X} \:.
\end{align}
It is important to notice that the parameters in Eq.~\eqref{EqUYUXbar} still depend on the geometric factor 
$g_{\mathrm{x}}$ and not on their respective geometric factors that could be derived analogously. 

To summarize, the above derivation constitutes
a self-consistent mapping between experimentally relevant recoil-energy scales and the dimensionless parameters used in the numerical simulation. By specifying the recoil energies $U_j^R$, the corresponding potential depths in system units $\tilde{U}_j$ are uniquely determined, while simultaneously fixing the dimensionless wavenumber $\tilde{k}$ through the harmonic reference frequency.

\end{document}